\documentclass[journal]{IEEEtran}
 \usepackage{amsmath,amssymb}
 \usepackage{subfigure}
 \usepackage{graphicx,graphics,color,psfrag}
 \usepackage{cite,balance}
 \usepackage{algorithm}
 \usepackage{accents}
 \usepackage{amsthm}
 \usepackage{bm}
 \usepackage{url}
 \usepackage{algorithmic}
 \usepackage[english]{babel}
 \usepackage{multirow}
 \usepackage{enumerate}
 \usepackage{cases}
 \usepackage{stfloats}
 \usepackage{dsfont}
 \usepackage{color,soul}
 \usepackage{amsfonts}
  \usepackage{tcolorbox}
\usepackage[utf8]{inputenc}
 \usepackage{cite,graphicx,amsmath,amssymb}
 \usepackage{subfigure}
 \usepackage{fancyhdr}
 \usepackage{hhline}
 \usepackage{graphicx,graphics}
 \usepackage{array,color}
 \usepackage{amsmath}
 \usepackage{amsthm}

\newtheorem{proposition}{Proposition}
\newtheorem{remark}{Remark}
\newtheorem{theorem}{Theorem}
\newtheorem{lemma}{Lemma}

\newtheorem{assumption}{Assumption}

\include{header}
\IEEEoverridecommandlockouts

\begin{document}

\title{Transmission Power Control for Over-the-Air Federated Averaging at Network Edge}
\author{Xiaowen Cao, Guangxu Zhu, Jie Xu, and Shuguang~Cui\\
\thanks{X. Cao is with the Future Network of Intelligence Institute (FNii), The Chinese University of Hong Kong (Shenzhen), Shenzhen, China, and the School of Information Engineering, Guangdong University of Technology, Guangzhou, China (e-mail: caoxwen@outlook.com).}
\thanks{G. Zhu is with Shenzhen Research Institute of Big Data, Shenzhen, China  (e-mail: gxzhu@sribd.cn). }
\thanks{J. Xu is with the School of Science and Engineering (SSE) and the FNii, The Chinese University of Hong Kong (Shenzhen), Shenzhen, China (e-mail: xujie@cuhk.edu.cn). J. Xu is the corresponding author.}
%\thanks{ Z. Wang is with China Academy of Information and Communications Technology, Beijing, China (e-mail: zhiqin.wang@caict.ac.cn). }
\thanks{S. Cui is with Shenzhen Research Institute of Big Data, Shenzhen, China, and the SSE and the FNii, The Chinese University of Hong Kong (Shenzhen), Shenzhen, China (e-mail: shuguangcui@cuhk.edu.cn). }
}

\markboth{}{}
\maketitle

\setlength\abovedisplayskip{2pt}
\setlength\belowdisplayskip{2pt}

%\vspace{-1cm}

\begin{abstract}%\vspace{-0.3cm}
%\emph{Federated edge learning} (FEEL) is expected to find abundant applications in \emph{beyond} 5G (B5G) and 6G networks towards the vision of connected intelligence, in which distributed edge devices need to cooperatively train machine learning models by using their local data while preserving privacy. Towards this end, \emph{over-the-air computation} (AirComp) has emerged as a new analog \emph{non-orthogonal multiple access} (NOMA) technique for model/gradient-updates aggregation in FEEL, which allows multiple edge devices to concurrently transmit their local model/gradient updates for ``one-shot'' aggregation at edge server, thus enhancing the communication efficiency of FEEL at the cost of aggregation errors caused by channel fading and noise.
\emph{Over-the-air computation} (AirComp) has emerged as a new analog power-domain \emph{non-orthogonal multiple access} (NOMA) technique for low-latency model/gradient-updates aggregation in \emph{federated edge learning} (FEEL).  By integrating communication and computation into a joint design, AirComp can significantly enhance the communication efficiency, but at the cost of aggregation errors caused by channel fading and noise.
This paper studies a particular type of FEEL with  \emph{federated averaging} (FedAvg) and AirComp-based model-update aggregation, namely \emph{over-the-air} FedAvg (Air-FedAvg). We investigate the transmission power control to combat against the AirComp aggregation errors for enhancing the training accuracy and accelerating the training speed of Air-FedAvg. Towards this end, we first analyze the convergence behavior (in terms of the optimality gap) of Air-FedAvg with aggregation errors at different outer iterations. Then, to enhance the training accuracy, we minimize the optimality gap by jointly optimizing the transmission power control at edge devices and the denoising factors at edge server, subject to a series of power constraints at individual edge devices.
Furthermore, to accelerate the training speed, we also minimize the training latency of Air-FedAvg with a given targeted optimality gap, in which learning hyper-parameters including the numbers of outer iterations and local training epochs are jointly optimized with the power control.
%By comparing the training latency of Air-FedAvg versus the conventional FedAvg with digital \emph{orthogonal multiple access}, dubbed as OMA-FedAvg, it shows the benefit of AirComp-based NOMA in FEEL.
Finally, numerical results show that the proposed transmission power control policy achieves significantly faster convergence for Air-FedAvg, as compared with benchmark policies with fixed power transmission or per-iteration \emph{mean squared error} (MSE) minimization. It is also shown that the Air-FedAvg achieves an order-of-magnitude shorter training latency than the conventional FedAvg with digital \emph{orthogonal multiple access} (OMA-FedAvg).
\end{abstract}

%
%\vspace{-0.8cm}
\begin{IEEEkeywords}%\vspace{-0.4cm}
Federated edge learning, federated averaging, over-the-air computation, non-orthogonal multiple access, power control.
\end{IEEEkeywords}

\vspace{-0.5cm}
%%\vspace{-0.2cm}
\section{Introduction}\label{sec:intro}

With the advancements in {\it artificial intelligence} (AI) and {\it Internet of Things} (IoT), cellular networks are experiencing a paradigm shift from connected everything in 5G to connected intelligence in {\it beyond 5G} (B5G) or 6G \cite{KLetaif19_6G}.  Recent trend has also witnessed the spreading of AI algorithms from the centralized cloud to the distributed network edge \cite{Survey_FEEl,Debbah19}, in order to efficiently utilize the distributed big data generated by the ever-increasing edge devices (such as sensors, IoT devices, and smartphones). Among various edge (distributed) machine learning approaches \cite{Ekram21_Survey}, {\it federated edge learning} (FEEL) is particularly appealing due to its privacy preserving feature, and thus has been envisioned as one of the candidate 6G techniques to enable abundant applications such as autonomous driving \cite{SPokhrel20_TCOM}, intelligent health \cite{XShen21_Health}, and industrial IoT \cite{DNguyen2021Ar}.

In particular, FEEL corresponds to implementing distributed {\it stochastic gradient descent} (SGD) over wireless networks, which allows distributed edge devices to collaboratively train a shared AI model by using their local data without compromising the privacy \cite{Konecny2016aa_FL,DLiu2020Ar}. The training process is to find optimized AI-model parameters by minimizing a properly designed loss function in an iterative manner. To be specific, at each outer iteration or communication round, the edge server first broadcasts the global AI-model parameters to edge devices, such that they can synchronize their local models; next, the edge devices update their respective local models or gradients by using their own local data, and then upload the updated local models/gradients back to the edge server, which aggregates them to update the global model.
 Note that due to the high dimensionality of each model/gradient update (usually comprising millions to billions of parameters), the frequent communication (for model/gradient uploading) and computation (for aggregation) from the edge devices to the edge server become the performance bottleneck for FEEL, especially when the number of edge devices sharing the same wireless medium becomes large. Therefore, to relieve such a bottleneck, it is important to design new multiple access communication techniques to facilitate the model uploading and aggregation from a large number of edge devices.

Conventionally, {\it orthogonal multiple access} (OMA) (e.g., {\it time division multiple access} (TDMA) and {\it orthogonal frequency-division multiple access} (OFDMA)) and power-domain {\it non-orthogonal multiple access} (NOMA) \cite{YLiu17_NOMA,ZDing17_Jsac,Liu2021_NGMA} have been implemented in FEEL as separated model uploading (communication) and aggregation (computation) approaches, in which the edge server first decodes each device's messages individually, and then aggregates them in a separate step \cite{Chen20FL,Mo2020aa,MChen21_TWC,Tran19_Infocom}. However, due to the separated communication and computation as well as individual message transmission and decoding, these conventional multiple access schemes may lead to significant transmission delays, thus limiting the training efficiency of FEEL.
To overcome this issue, {\it over-the-air computation} (AirComp) \cite{nomo_function_Nazer,Gastpar08,Zhu2021ComMag,Cao_PowerTWC} has emerged as a new analog NOMA technique.
 In contrast to the existing digital NOMA schemes that aim to decode multiple users' individual messages by mitigating the harmful inter-user interference via {\it successive interference cancellation} (SIC) \cite{YLiu17_NOMA,ZDing17_Jsac,Liu2021_NGMA}, the analog NOMA or AirComp aims to compute functions based on multiple users' individual data, in which the inter-user interference is harnessed as a beneficial factor for computation.
 Due to the benefit introduced by the communication-computation integration, AirComp has found successful applications in FEEL to motivate the so-called {\it over-the-air} FEEL (Air-FEEL), in which multiple edge devices are allowed to transmit their local model/gradient updates concurrently, such that the edge server can exploit the AirComp for ``one-shot" aggregation. As compared to conventional FEEL with digital OMA or NOMA, the Air-FEEL is expected to significantly reduce the communication and computation latencies for  model uploading and aggregation, thus enhancing the training efficiency.

%Due to the frequent model/gradient updates exchange between the edge server and possibly massive number of edge devices in FEEL, the resultant wireless communication overheads become the performance bottleneck hindering the large-scale implementation of FEEL.

%Depending on different averaging approaches based on models and gradients,
In general, the Air-FEEL can be classified into two categories \cite{FedAvg}, namely {\it over-the-air  federated stochastic gradient descent} (Air-FedSGD) and  {\it over-the-air federated averaging} (Air-FedAvg), which feature gradient and model aggregations at each outer iterations, respectively. %, which updates the global models based on gradient averaging and model averaging,  respectively.
In the literature, several prior works \cite{ Amiri2020TSP,MAmiri2020TWC,GZhu2020Ar,RJiang2020Ar,HG21IoT,JZhang2021Ar,Ni2021_NOMA1,NZhang2020Ar,Cao2021AirFEEL} investigated the Air-FedSGD from different perspectives.
%Air-FedSGD features {\it gradient aggregation}, in which local gradient estimates at edge devices are conveyed and aggregated for global model updating at the edge server \cite{ Amiri2020TSP,MAmiri2020TWC,GZhu2020Ar,RJiang2020Ar,HG21IoT,JZhang2021Ar,Ni2021_NOMA1,NZhang2020Ar,Cao2021AirFEEL}, while Air-FedAvg features {\it model aggregation}, where local models instead of the gradients are uploaded and aggregated \cite{GZhu2020TWC,XFan2021Ar04,SXia2020Ar,KYang2020TWC,SWang2021Ar,Ni2021_NOMA_RIS}.
For instance, the authors in \cite{Amiri2020TSP,MAmiri2020TWC} used a source-coding algorithm to compress the gradient updates by exploiting their sparsity.
\cite{GZhu2020Ar} proposed a digital Air-FedSGD solution to make Air-FedSGD compatible with the modern digital modulation. By extending the work in \cite{GZhu2020Ar}, \cite{RJiang2020Ar} studied a design with  one-bit quantization and modulation at the edge devices and majority-vote based decoding at the edge server.
 Furthermore, the authors in \cite{HG21IoT,JZhang2021Ar} proposed to optimize the hyper-parameters (such as learning rates)  for accelerating the training process, while the authors in \cite{Ni2021_NOMA1,NZhang2020Ar,Cao2021AirFEEL} investigated the transmission power control policies to reduce the computation distortion over wireless transmission.
%Although the gradient aggreagtion approach have a better convergence guarantee, it may cost heavy communication due to the frequent updates at every outer iteration.
Despite the research progress, however, the Air-FedSGD may lead to heavy communication costs, as frequent gradient aggregations are needed after every single local gradient update or local training epoch.

As an alternative approach,  Air-FedAvg with model aggregations can run multiple local training epochs before each global model aggregation, but requires much less frequent model aggregation, thus leading to lower communication costs than the Air-FedSGD counterpart.
This has triggered an increasing number of researchers to study Air-FedAvg under different system setups \cite{GZhu2020TWC,XFan2021Ar04,SXia2020Ar,KYang2020TWC,SWang2021Ar,Ni2021_NOMA_RIS}.
%To further improve the communication efficiency, accessing device scheduling is a potential approach.
For instance, the authors in \cite{GZhu2020TWC} considered a broadband Air-FedAvg system, for which a set of interesting communication-learning tradeoffs was derived to guide the device scheduling.
Subsequently,  the joint design of device scheduling and channel-inversion based power scaling was investigated in \cite{XFan2021Ar04} and a {\it channel state information} (CSI) based device selection scheme was proposed in \cite{SXia2020Ar} to achieve reliable model aggregation.
Then, by considering multi-antenna Air-FedAvg systems, a joint device scheduling and receive beamforming design was studied in \cite{KYang2020TWC}, and a unit-modulus analog receive beamforming design was proposed in \cite{SWang2021Ar}.
Despite the progress made in prior work, however, how to analytically characterize the effect of the AirComp induced model aggregation errors on the learning performance of Air-FedAvg, and how to quantify the practical performance gain of Air-FedAvg over conventional OMA-FedAvg are still open problems that have not been well understood. This thus motivates the current work in this paper.

%\subsection{Main Contributions}%\vspace{-0.1cm}

This paper studies an Air-FedAvg system, which consists of multiple edge devices and one edge server. By considering smooth learning models satisfying the Polyak-{\L}ojasiewicz inequality, a learning performance metric, namely the optimality gap, is derived to link the convergence performance with the AirComp aggregation errors over outer iterations.
%exploit the fact that the aggregation distortion in different communication rounds may have distinct impact on the learning performance .
%{\color{blue}To the best of the authors' knowledge, this is the first attempt to mathematically characterize the impact of AirComp error on the Air-FedAvg performance.}
Based on the analytical results, we propose an optimized power control policy for directly minimizing the optimality gap. Moreover, we further compare training latency between the Air-FedAvg and the conventional OMA-FedAvg to quantify the performance gain achieved by AirComp in FEEL.
The main contributions are elaborated as follows.
\begin{itemize}
	\item { \bf Convergence analysis for Air-FedAvg:}
	First, we analyze the optimality gap of the loss function over different outer iterations, to capture the impact of aggregation errors (i.e., the bias and {\it mean squared error} (MSE) of the model aggregation) on the convergence performance of the Air-FedAvg algorithm. It is revealed that within finite number of outer iterations, the aggregation errors at later iterations contribute more (with higher weights) to the optimality gap than those at earlier iterations. To our best knowledge, this is the first attempt to mathematically characterize the impact of AirComp error on the Air-FedAvg performance.
	\item { \bf Optimality gap minimization via power control:}	
	Building on the convergence analysis, we formulate an optimality gap minimization problem for Air-FedAvg, by jointly optimizing the transmission power control at edge devices and the denoising factors at edge server, subject to a set of power constraints at individual edge devices. Due to the coupling of the power control and denoising factors, the formulated problem is non-convex and thus difficult to be solved optimally. To tackle this difficulty, we propose an efficient algorithm based on the  alternating optimization to obtain a high-quality solution.
%	we proposed an alternating optimization based algorithm to solve the formulated problem, by first optimizing the denoising factor under any given transmission power scaling factor, and then optimizing the power control.
%	Fortunately, by transformed into convex forms via a variable replacement,  the formulated problem can be optimally solved by the Lagrangian duality method. The optimized power control solutions establish a regularized channel inversion structure, where the regularization term at each edge device is related to all other devices' average power budgets.
	\item {  \bf Training latency minimization via joint power control and hyper-parameter optimization:} Next, to accelerate the training speed, we further consider the training latency minimization for Air-FedAvg with a given targeted optimality gap, in which the learning hyper-parameters including the numbers of outer iterations and local training epochs are optimized  jointly with transmission power control. Based on the results, we quantify the training latency gain by Air-FedAvg over OMA-FedAvg to show the benefit of the new AirComp-enabled NOMA approach in enhancing the FEEL efficiency.
%	 For OMA-FedAvg, we adopt a {\it time division multiple access} (TDMA) protocol to schedule the model uploading. Thus, due to the interference-free model transmission,  the aggregation error is specified as the quantization error of the model parameters, and thus its derived optimality gap is related to the quantization level.
%	 For fair comparison, we formulate latency minimization problems for both Air-FedAvg and OMA-FedAvg with a given targeted optimality gap  by jointly optimizing the resource allocation, number of outer iterations and local training epoch. Both problems are non-convex and hard to solved optimally due to the coupling between variables. For the latency minimization problem of Air-FedAvg, once given the number of outer iterations and local training epochs, it can reduce to the optimality gap minimization problem via power control. Hence, based on alternating optimization technique, we propose an efficient algorithm to obtain a high-quality suboptimal solution.
%	 As for OMA-FedAvg, we first obtain the optimal number of outer iterations and local training epochs via a bisection search on the optimality constraint, and then leverage the Lagrangian duality to obtain the optimal transmission power scaling factor and time allocation policy.
   \item  {\bf Performance evaluation:} Finally, we conduct extensive simulations on both synthetic and real datasets to evaluate the performance of the optimized power control for Air-FedAvg. %by considering the ridge regression with synthetic dataset, and handwritten digit recognition using MNIST dataset with a {\it convolution neural network} (CNN).
   It is shown that the proposed power control policy achieves significantly faster convergence rate (or lower optimality gap) than the benchmarking schemes with fixed power transmission and per-iteration MSE minimization. This is due to the fact that the proposed policy can better handle the aggregation errors over iterations based on their contributions to the optimality gap.
  It is also shown that the Air-FedAvg achieves an order-of-magnitude shorter training latency than the conventional OMA-FedAvg.
%   It is also shown that the latency minimization performance of Air-FedAvg outperforms that of OMA-FedAvg. Although the required number of global updates in Air-FedAvg is larger than that in OMA-FedAvg, Air-FedAvg can achieve significant latency reduction up to 10 times due to the fact that the per-iteration latency of OMA-FedAvg is higher and scaling with number of accessing devices.
   \end{itemize}

The remainder of this paper is organized as follows. Section II presents the system model of Air-FedAvg. Section III analyzes the optimality gap in the presence of AirComp aggregation errors. Section IV minimizes the optimality gap by optimizing the transmission power control. Section V compares the minimally achievable training latency under targeted optimality gap of Air-FedAvg versus that of OMA-FedAvg. Finally, Section VI provides simulation results, followed by the conclusion in Section VII.

%{\color{red} The contribution part would be added ASAP

%{\it Notations:}  Bold lowercase letters refer to column vectors.
%$\mathbb{E}(\cdot)$ denotes the expectation operation; the superscript $\dagger $ represents the transpose operation; $\nabla$ is the gradient operator, and $(x)^+\triangleq\max\{0,x\}$.
%For a set $\mathcal{A}$, $|\mathcal{A}|$ denotes its cardinality.
%$\|\bm a\|$ denotes the Euclidean norm of vector $\bm a$. $\bf I$ denotes the identity matrix.

\section{System Model}\label{sec:system}
\begin{figure}
\centering
 \setlength{\abovecaptionskip}{-4mm}
\setlength{\belowcaptionskip}{-4mm}
    \includegraphics[width=3.5in]{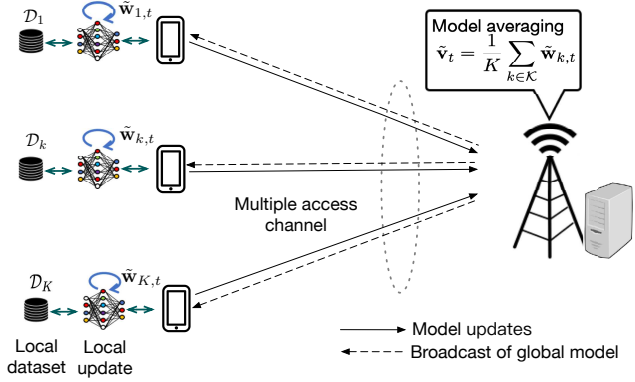}
\caption{Illustration of the Air-FedAvg system with AirComp-based model aggregation. } \label{fig:model}
\vspace{-0.4cm}
\end{figure}

We consider an Air-FedAvg system as shown in Fig.~\ref{fig:model}, in which $K$ edge devices are coordinated by one edge server to train a shared machine learning model via the AirComp-based model aggregation. In the following, we introduce the FedAvg algorithm and the AirComp-based model aggregation, respectively. % as will be elaborated next.
%In this paper, we consider two approaches including orthogonal and non-orthogonal multiple access to implement the model aggregation.

\subsection{Federated Averaging at Network Edge}

Suppose that the machine learning model to be trained is represented by the parameter vector ${\bf w}\in\mathbb{R}^q$, where ${\bf w}=[w_1,\cdots,w_q]^{ \dagger}$ with $q$ denoting the model size and the superscript $\dagger$ denoting the transpose operation.
Let ${\mathcal D}_k$ denote the local dataset at edge device $k\in \mathcal{K} \triangleq \{1,...,K\}$.
 Define $f({\bf w},{\bf \xi}_i)$ as the sample-wise loss function quantifying the prediction error of model $\bf w$ on a sample ${\bf \xi}_i$.
Then the local loss function of the model vector $\bf w$ on ${\mathcal D}_k$ is
\begin{align}\label{LocalLossFunction}
\tilde{F}_k({\bf w})=\frac{1}{|{\mathcal D}_k|} \sum \limits_{{\bf \xi}_i\in{\mathcal D}_k} f({\bf w},{\bf \xi}_i),
\end{align}
where $|{\mathcal D}_k| $ denotes the cardinality of ${\mathcal D}_k$.
Assume that the dataset at each device has the same size with $|{\mathcal D}_k|=\bar{D}, \forall k\in\mathcal{K}$.
Accordingly, the global loss function on all $K$ devices'  datasets ${\mathcal D}=\cup_{k\in\mathcal K} {\mathcal D}_k$ evaluated on parameter vector $\bf w$ is given by
\begin{align}\label{GlobalLossFunction}
F({\bf w})=\sum\limits_{k\in\mathcal{K}} \frac{|{\mathcal D}_k|}{|{\mathcal D}| } F_k({\bf w})=\sum\limits_{k\in\mathcal K} \frac{\bar{D}}{ K\bar{D}} {F}_k({\bf w})=\frac{1}{ K} \sum\limits_{k\in\mathcal K} {F}_k({\bf w}).
\end{align}

Mathematically, the objective of the training process is to minimize the global loss function $F({\bf w})$ in \eqref{GlobalLossFunction} by optimizing the model parameter $\bf w$:
 \begin{align}\label{OptimalParameter}
 {\bf w}^{\star}=\arg \min_ {\bf w} F({\bf w}).
\end{align}
To solve problem \eqref{OptimalParameter} while preserving the data privacy for each device, in this paper, we consider the FedAvg algorithm, which is implemented iteratively in a distributed manner as follows.

Let $T$ denote the number of outer iterations with $\mathcal{T} \triangleq\{1,\cdots,T\}$. At each outer iteration $t\in \mathcal T$, the edge server broadcasts the global machine learning model $\tilde{\bf v}_{t}$ to the $K$ devices.
After that, each device $k$ first updates the local machine learning model as ${\bf w}_{k,t,0}=\tilde{\bf v}_{t}$, and then performs a total of  $\Omega$ local training epochs, each based on a randomly sampled mini-batch from the local dataset denoted by $ \tilde{\mathcal D}_k$, with size of $n_b = | \tilde{\mathcal D}_k|, \forall k\in \mathcal K$. Let $\tilde{F}_k({\bf w}_{k,t,\ell},\tilde{\mathcal D}_k )$ define the local gradient estimate over $ \tilde{\mathcal D}_k$ at $\ell$-th local training epoch, and thus we have
\begin{align}
	{\bf w}_{k,t,\ell+1} ={\bf w}_{k,t,\ell}-\gamma_t \nabla \tilde{F}_k({\bf w}_{k,t,\ell},\tilde{\mathcal D}_k ),\forall \ell\in \{0,...,\Omega-1\},
\end{align}
 where $\gamma_t$ denotes the learning rate at outer iteration $t$.
Then each device $k$ uploads the local model updates $\tilde{\bf w}_{k,t+1}={\bf w}_{k,t,\Omega}$ to the edge server.
 After collecting $K$ devices' updated models $\{\tilde{\bf w}_{k,t+1}\}$, the edge server aggregates them to obtain the updated global model as\footnote{Although we consider the same dataset size at  different edge devices ($|\mathcal{D}_k|= \bar D, \forall k\in \mathcal K$), our proposed Air-FedAvg designs in this paper can be easily extended to the case when $|\mathcal{D}_k|$'s are different. In this case, we only need to revise the global model parameter in \eqref{Sys_Globalmodel} as a weighted-average of the local ones, i.e., $\tilde{\bf v}_{t+1}=\sum\limits_{k\in\mathcal K}\frac{|\mathcal{D}_k|}{|\mathcal{D}|}\tilde{\bf w}_{k,t+1}$. Based on  AirComp, the desired weighted aggregation of the local gradient estimate can be  easily obtained by adding an additional pre-processing $\psi(\cdot)$ on the transmitted signal $s_k$ with $\psi(s_k)=\sum\limits_{k\in\mathcal K}\frac{|\mathcal{D}_k|}{|\mathcal{D}|}s_k$.}
 \begin{align}\label{Sys_Globalmodel}
 	\tilde{\bf v}_{t+1}\!=\!\frac{1}{ K}\! \sum\limits_{k\in\mathcal K} \!\!\tilde{\bf w}_{k,t+1}.
 \end{align}
The process repeats until the number of outer iterations $T$ is met.

%}

%\vspace{-0.45cm}
\subsection{Model Aggregation via Over-the-air Computation}\label{CommunicationModel}
%\vspace{-0.2cm}

In the FedAvg, the communication overhead for model uploading and aggregation becomes the performance bottleneck,  especially when the number of edge devices becomes large.
To accelerate the model aggregation process,  we consider the AirComp-enabled analog NOMA approach to enable integrated communication and computation for multiple devices.
%for accelerating the aggregation process.

For the purpose of illustration, we consider a frequency non-selective block fading channel model, where the wireless channels remain unchanged over each outer iteration, and may change over different iterations.
% For simplicity, each outer iteration consists of $q$ time slots to enable the uncoded transmission of a full model vector.
Besides, each edge device is assumed to perfectly know its own CSI, so that they can compensate for the phases introduced by the wireless channels, and the edge server is assumed to know the global CSI to facilitate the power control.

%\subsubsection{\bf AirComp Scheme}
Let $\hat h_{k,t}$ denote the complex channel coefficient from device $k$ to the edge server at outer iteration $t$, and $h_{k,t}$ denote its magnitude with $h_{k,t}=|\hat h_{k,t}|$, $\forall k\in\mathcal{K}, t\in\mathcal{T}$.
For model-uploading phase at outer iteration $t$, let $\tilde{\bf w}_{k,t}\in \mathbb R^{q}$ denote the transmitted message by each device $k$ by using $q$ symbols, and $p_{k,t}$ denote the transmission power scaling factor. By allowing all devices to transmit simultaneously with proper phase control, the received  signal (after phase compensation) at the edge server  is given by
 \begin{align}\label{sys_ReceivedSignal}
 	{\bf y}_{t}=\sum\limits_{k\in\mathcal K}h_{k,t}\sqrt{p_{k,t}}\tilde{\bf w}_{k,t}+{\bf z}_{t}, \forall t\in\mathcal{T},
 \end{align}
 in which ${\bf z}_{t}\in\mathbb{R}^q$ denotes the {\it additive white Gaussian noise} (AWGN) with ${\bf z}_{t}\sim{\mathcal CN}(0,\sigma_z^2\bf I)$, with $\sigma_z^2$ denoting the noise power and $\bf I$ denoting an identity matrix.
Based on \eqref{sys_ReceivedSignal}, the edge server estimates the global model as ${\bf v}_{t}$ by implementing a denoising factor $\eta_t$, i.e.,
\begin{align}\label{sys_ComGlobalGradient}
{\bf v}_{t}=\frac{{\bf y}_{t}}{\sqrt{\eta_{t}}K}=\frac{\sum\limits_{k\in\mathcal K}h_{k,t}\sqrt{p_{k,t}}\tilde{\bf w}_{k,t}+{\bf z}_{t}}{\sqrt{\eta_{t}}K} , \forall t\in\mathcal{T}.
\end{align}
%where $\eta_t$ denotes the denoising factor.
%{\color{blue}
In this case,  the communication error caused by the AirComp-based aggregation {\it with respect to} (w.r.t.) the global gradient estimation $\tilde{\bf v}_{t}$ is given by
\begin{align}\label{sys_Err}
{\bm \varepsilon}_{t}={\bf v}_{t}-\tilde{\bf v}_{t},  t\in\mathcal{T}.
\end{align}
%where $\tilde{\bf v}_{t}\triangleq \frac{1}{K}\sum\limits_{k\in\mathcal{K}} \tilde{\bf w}_{k,t}$.
Suppose that each device $k\in \mathcal K$ is subject to a maximum power budget $P^{\rm max}_k$ at each outer iteration $t$ and a maximum average transmit power budget $P^{\rm ave}_k$ across all outer iterations. We then have
\begin{align}
\frac{1}{q} ~p_{k,t} \mathbb{E}\left(\left\|\tilde{\bf w}_{k,t}\right\|^2\right) \leq P^{\rm max}_k,~\forall k\in{\mathcal K}, ~ t\in\mathcal{T},\label{sys_bar_P_max1}\\
\frac{1}{qT}\sum \limits_{t\in\mathcal{T}}p_{k,t} \mathbb{E}\left(\left\|\tilde{\bf w}_{k,t}\right\|^2\right) \leq P^{\rm ave}_k,~\forall k\in{\mathcal K},\label{sys_bar_P_ave1}
\end{align}
where $\mathbb{E}(\cdot)$ denotes the statistical expectation, and $\| \cdot\|$ denotes the Euclidean norm of  a vector.
Note that $P^{\rm max}_k$ should be no smaller than $P^{\rm ave}_k$ in order for both constraints to be valid.

 \section{Convergence Analysis for Air-FedAvg}

In order to facilitate the Air-FedAvg design later, this section presents a convergence analysis framework for Air-FedAvg subject to aggregation errors, in which the optimality gap is adopted as the performance metric.
% The convergence results will be used to facilitate the Air-FedAvg design later.
\subsection{Assumptions on Loss Functions and Gradient Estimates}
For the purpose of analysis,  we make several assumptions on the loss function and gradient estimates, as commonly adopted in the literature  \cite{Li2019_Noniid,SWang19_FedAvg}.

 \begin{assumption}[Lipschitz-Continuous Gradient]\label{Assump_Smooth}\emph{
The gradients of loss functions $F({\bf w})$ and $F_k({\bf w})$ are Lipschitz continuous with a common non-negative constant $L>0$, i.e., for any ${\bf w}, {\bf v} \in\mathbb{R}^q$ and $k\in\mathcal{K}$, it holds that
\begin{align}
\|\nabla	 F({\bf w})-\nabla F({\bf v})\|\leq L\|{\bf w}- {\bf v}\|,\\
\|\nabla	 F_k({\bf w})-\nabla F_k({\bf v})\|\leq L\|{\bf w}- {\bf v}\|,\label{Smooth_gra_k}
\end{align}
where $\nabla F({\bf w})$ and $\nabla F_k({\bf w})$ denote the gradients of the loss function evaluated at point ${\bf w} \in\mathbb{R}^q$ at the edge server and device $k\in\mathcal{K}$, respectively.
As a consequence, for any ${\bf w}, {\bf v} \in\mathbb{R}^q$, we have
%following which, there also exists an important consequence given by
\begin{align}
F({\bf w}) &\le  F({\bf v}) + \nabla F({\bf v})^T (\!{\bf w}\!\!-\! {\bf v})+\frac{L}{2}\|{ \bf w}-{\bf v}\|^2, \label{smooth_Func}\\
F_k({\bf w})& \le  F_k({\bf v}) + \nabla F_k({\bf v})^T (\!{\bf w}\!\!-\! {\bf v})+\frac{L}{2}\|{ \bf w}-{\bf v}\|^2.\label{smooth_Func_k}
%\sum_{i=1}^{q}\! \!L_i({{ w}_i-\!{w}^{\prime}_i})^2\!, \forall {\bf w}, {\bf w}^{\prime},
\end{align}
}
\end{assumption}
\begin{assumption}[Polyak-{\L}ojasiewicz Inequality]\label{Assump_PL}\emph{
Let $F^{\star}$ denote the optimal loss function value for problem \eqref{OptimalParameter}. There always exists a constant $\mu\ge 0$ such that any global loss function $F({\bf w})$ satisfies the following Polyak-{\L}ojasiewicz (PL) condition:
\begin{align*}
	\| \nabla F({\bf w}) \|_2^2 \ge 2\mu(F({\bf w})-F^{\star}).
\end{align*}
%where $\delta\ge 0$ is a constant.
}
\end{assumption}
 \begin{assumption}[Variance Bound]\label{Assum_VarianceBound}\emph{The variance of stochastic gradients $\nabla \tilde{F}_k({\bf w},\tilde{\mathcal D}_k)$ at each device $k\in\mathcal{K}$ w.r.t. a set of randomly selected samples $\tilde{\mathcal D}_k$  is bounded by
\begin{align}
	&\mathbb{E}\left(\left\| \nabla F_k({\bf w} )-\nabla \tilde{F}_k({\bf w},\tilde{\mathcal D}_k)\right\|^2\right)\le \frac{\hat{\phi}_k^2}{n_b}, ~\forall k\in\mathcal K,
\end{align}
where $\hat{\bm \phi}=[\hat{\phi}_1,\cdots,\hat{\phi}_K]^{\dagger}$ is a vector of non-negative constants and $n_b$ is the mini-batch size.
 }
\end{assumption}
 \begin{assumption}[Gradient Bound]\label{Assum_GraBound}\emph{The expected squared norm of stochastic gradient $\nabla \tilde{F}_k({\bf w},\tilde{\mathcal D}_k)$ at each device $k\in\mathcal{K}$ is bounded by a positive constant $G_k$, i.e.,
\begin{align}
	&\mathbb{E}\left(\left\| \nabla \tilde{F}_k({\bf w},\tilde{\mathcal D}_k) \right\|^2\right)\le G_k^2, ~\forall k\in\mathcal K.
\end{align}
%where $G_k, \forall k\in\mathcal{K}$ is a positive constant.
 }
\end{assumption}

%To characterize the divergence between the local gradient and the global gradient, we define a metric named {\it Gradient Divergence}, which can serve as a rough measure at the data heterogeneity over edge devices.

 \begin{assumption}[Gradient Divergence]\label{Assum_GraDiv}\emph{
 Let $\delta_k\geq 0$ denote the upper bound of the gradient divergence for edge device $k$. It must hold that
\begin{align}
	&\left\| \nabla F_k({\bf w}) - \nabla F({\bf w}) \right\|\le \delta_k, ~\forall k\in\mathcal K.
\end{align}
 }
\end{assumption}

 \begin{assumption}[Model Bound]\label{Assum_ModBou}\emph{
 Let $W_k\geq 0$ denote the upper bound of the model parameter for edge device $k$, i.e., it  holds that
\begin{align}
	\mathbb{E}\left(\left\|\tilde{\bf w}_{k,t}\right\|^2\right)\leq W_k^2,~ \forall k\in\mathcal{K}.
\end{align}
 }
\end{assumption}

Notice that typical functions satisfying Assumptions \ref{Assump_Smooth} and \ref{Assump_PL} include logistic regression, linear regression, and least squares.
Also note that the gradient divergence in Assumption \ref{Assum_GraDiv} characterizes the divergence between the local gradient and the global gradient, which serves as a rough measure of the data heterogeneity over edge devices.

\subsection{Optimality Gap Analysis }
Now, based on Assumptions \ref{Assump_Smooth}-\ref{Assum_GraDiv}, we analyze the convergence performance of FedAvg in terms of the optimality gap.
First, we analyze the optimality gap under given communication (or model aggregation) errors ${\bm \varepsilon}_{t}$'s, as given in \eqref{sys_Err} for outer iterations.
After each outer iteration $t$, the parameter vector becomes ${\bf v}_{t+1}$, and the corresponding optimality gap is defined as $F\left({\bf v}_{t+1}\right)-F^{\star}$.
%the expected mean and mean square of ${\bm \varepsilon}_{t}$
Let $\mathbb{E}\left({\bm \varepsilon}_{t}\right)$ and $\mathbb{E}\left(\|{\bm \varepsilon}_{t}\|^2\right)$ denote the bias and MSE of the global model aggregation at each outer iteration $t\in\mathcal{T}$, respectively.
Then, by considering a properly chosen decaying learning rate, we establish the following theorem, which characterizes the optimality gap w.r.t. the communication errors.

\begin{theorem}%[Impact of aggregation error on Convergence]
\label{Theo_OG_NOMA}\emph{Suppose that the FedAvg algorithm is implemented with diminishing learning rates $\gamma_t=\frac{\beta}{t+a}, \forall t\in\mathcal{T}$ \cite{Bottou2018}, where $a>0$ and $\beta\geq \frac{1}{\mu(\Omega-1)}$, such that $\gamma_1\leq\frac{1}{L\Omega}$.
Then, the expected optimality gap satisfies the following inequality:
\begin{align}\label{OG_NOMA}
&\mathbb{E}\left(F\left({\bf v}_{T}\right)\right)-F^{\star}\notag\\
&\leq \prod_{t\in\mathcal{T}}C_t \left(F\left({\bf v}_{0}\right)-F^{\star}\right)  +\sum_{t=1}^{T} J_t \left( \gamma_{t-1} \Omega B+  \gamma_{t-1}^2 \Omega^2 V\right) +\notag\\
& \sum_{t=1}^{T}\!\frac{J_t}{2}\left(\frac{1}{\gamma_{t-1} }\left\|\mathbb{E}\left({\bm \varepsilon}_{t}\right)\right\|^2\!+\!\left({L^2\gamma_{t-1} \Omega}\!+\!L\right)\mathbb{E}\left(\left\|{\bm \varepsilon}_{t}\right\|^2\right)\!\! \right)\!,\!
\end{align}
where $B= \frac{1}{K}\sum\limits_{k\in\mathcal{K}} \frac{\delta_k^2 +\phi_k^2}{2}$ captures the gradient variance bound and the gradient divergence with $\phi_k^2=\frac{\hat{\phi}_k^2}{ n_b} $, $ J_t\triangleq\frac{\prod_{i=t}^{T}C_t}{C_{t}}, \forall k\in\mathcal{K},\forall t\in\mathcal{T} $ with $C_t=1-(\Omega-1)\mu\gamma_t $,  and $V=  \frac{L }{ K} \sum\limits_{k\in\mathcal K}G_k^2$ captures the gradient bounds.
}
\end{theorem}
\begin{IEEEproof}
See Appendix~\ref{Proof_Theo_OG_NOMA}.
\end{IEEEproof}
\begin{remark}\label{Remark_Theorem1}\emph{
From Theorem \ref{Theo_OG_NOMA}, it is observed that the optimality gap in \eqref{OG_NOMA} critically depends on the learning rates $\{\gamma_t\}$ and the local training epochs $\Omega$, which are important hyper-parameters to be optimized for speeding up the convergence of training. It is also observed that as the coefficients $J_t$'s are monotonically increasing over $t$, and thus the bias  $\mathbb{E}\left({\bm \varepsilon}_{t}\right)$ and MSE $\mathbb{E}\left(\left\|{\bm \varepsilon}_{t}\right\|^2\right)$ at the later iterations (with larger $t$) generally contribute more on the optimality gap in \eqref{OG_NOMA}.
}
\end{remark}

% \subsection{Optimality Gap versus Aggregation Error  }\label{Sec_AggError}
%\subsection{Convergence Analysis of Air-FedAvg}

Next, we characterize the optimality gap of Air-FedAvg w.r.t. the transmission power control variables $\{p_{k,t}\}$ and denoising factors $\{\eta_t\}$, by specifying aggregation errors ${\bm \varepsilon}_{t}$'s as functions of $\{p_{k,t}\}$ and $\{\eta_t\}$.
%According to the definition of $ {\bm \varepsilon}_t$ in expression
Notice that based on \eqref{sys_Err}, the squared bias and MSE of model estimates at each outer iteration $t$ are obtained as
\begin{align}
\!\!\!\!\left\|\mathbb{E}\left({\bm \varepsilon}_{t}\right)\right\|^2\!&=\left\|\frac{1}{K}\sum\limits_{k\in\mathcal{K}}  \left(\frac{h_{k,t}\sqrt{p_{k,t}}}{\sqrt{\eta_t}}-1\right)\mathbb{E}\left(\tilde{\bf w}_{k,t}\right)\right\|^2\notag\\
&\leq\frac{1}{K}\sum\limits_{k\in\mathcal{K}}  \left(\frac{h_{k,t}\sqrt{p_{k,t}}}{\sqrt{\eta_t}}-1\right)^2W_k^2 ,\label{equ_bias_p}\\
\!\!\!\!\mathbb{E}\!\left(\left\|{\bm \varepsilon}_{t}\right\|^2\right)\!&=\!\mathbb{E}\!\left\|\frac{1}{K}\sum\limits_{k\in\mathcal{K}} \! \left(\!\frac{h_{k,t}\sqrt{p_{k,t}}}{\sqrt{\eta_t}}-\!1\!\right)\tilde{\bf w}_{k,t}\right\|^2\!\!\!+\!\frac{\sigma_z^2q}{\eta_tK^2}\notag\\
&\leq\frac{1}{K}\sum\limits_{k\in\mathcal{K}}  	\!\left(\frac{h_{k,t}\sqrt{p_{k,t}}}{\sqrt{\eta_t}}-\!1\!\right)^2\!\!W_k^2\!+\!\frac{\sigma_z^2q}{\eta_tK^2},\label{equ_unbias_p}\!
\end{align}
respectively, where the inequalities in \eqref{equ_bias_p}  and \eqref{equ_unbias_p} result from the combined use of Assumptions~\ref{Assum_VarianceBound} and \ref{Assum_ModBou}, together with the Cauchy's inequality.
Then based on Theorem \ref{Theo_OG_NOMA}, and by substituting \eqref{equ_bias_p} and \eqref{equ_unbias_p} into \eqref{OG_NOMA}, we have the following proposition.
%[Impact of aggregation error on the convergence of Air-FedAvg]
\begin{proposition}\label{Prop_OG_NOMA}\emph{By considering the diminishing learning rates $\gamma_t=\frac{\beta}{t+a}, \forall t\in\mathcal{T}$, with $a>0$ and $\beta\geq \frac{1}{\mu(\Omega-1)}$, such that $\gamma_1\leq\frac{1}{L\Omega}$, the expected optimality gap of Air-FedAvg satisfies that
%Then, we have
\begin{align}\label{Power_OG_NOMA}
\!\!\!&\mathbb{E}\left(F\left({\bf v}_{T}\right)\right)-F^{\star}\leq  \Phi\left(\{p_{k,t}\},\{\eta_t\},T, \Omega \right)\triangleq \notag\\
&\prod_{t\in\mathcal{T}}C_t \left(F\left({\bf v}_{0}\right)-F^{\star}\right)  +\sum_{t=1}^{T}\!\left(\gamma_{t-1}\Omega B+  \gamma_{t-1}^2 \Omega^2 V\right)+ \notag\\
\!\!&\sum_{t=1}^{T}\!\frac{J_t(L\!+\!\gamma_{t-1}L^2\Omega)}{2}\!\left(\!\frac{1}{K}\sum\limits_{k\in\mathcal{K}} \! \left(\!\frac{h_{k,t}\sqrt{p_{k,t}}}{\sqrt{\eta_t}}\!-\!1\!\right)^2 \!\!W_k^2\!+\!\frac{\sigma_z^2q}{\eta_tK^2}\!\right)\notag\\
&+ \sum_{t=1}^{T}\frac{J_t}{2\gamma_{t-1}}\frac{1}{K}\sum\limits_{k\in\mathcal{K}}  \left(\frac{h_{k,t}\sqrt{p_{k,t}}}{\sqrt{\eta_t}}-1\right)^2W_k^2.
\end{align}
}
\end{proposition}
%\Phi\left(\{p_{k,t}\},\{\eta_t\},T, \Omega \right)\triangleq\notag\\
%	&~~
%[30,  Corollary 1]
%\begin{remark}\label{Remark_Prop1}\emph{
By comparing the optimality gap for Air-FedAvg in Proposition \ref{Prop_OG_NOMA} versus that for Air-FedSGD in \cite[Corollary 1]{Cao2021AirFEEL}, we observe the following essential differences.
First, while in the Air-FedSGD, the learning rates $\{\gamma_t\}$ can be reused to play the role of denoising factors for AirComp, in Air-FedAvg dedicated denoising factors $\{\eta_t\}$ should be used in order to suppress the AirComp signal-misalignment error  (i.e., the second term at the right-hand-side of \eqref{Power_OG_NOMA}). This makes the power control design problem in Air-FedAvg more challenging than that in Air-FedSGD \cite{Cao2021AirFEEL}, as will be shown in Section \ref{Sec_Opt}.
Next, as compared to Air-FedSGD with only one training epoch for each outer iteration, in Air-FedAvg the number of local training epochs $\Omega$  becomes a new design parameter affecting the optimality gap, thus making the optimality gap formula more complex than that for Air-FedSGD. In particular, as $\Omega$ increases, the coefficient $C_t$ in \eqref{Power_OG_NOMA} will decrease but the terms related to gradient divergence and gradient bound ($B$ and $V$) will increase. Therefore, $\Omega$ needs to be judiciously designed in Air-FedAvg for minimizing the optimality gap.
%would be enhanced while the discounting factor $J_t$ would be decreasing as $C_t$ decreases.
%}
%\end{remark}

%the learning rate can served as a scaling factor at the receiver like the role of  dedicated scaling factors in conventional AirComp to balance the misalignment error and noise perturbation, which cannot work on Air-FedAvg.

\section{Optimality Gap Minimization via Power Control}\label{Sec_Opt}

%Based on the derived optimality gap in Proposition~\ref{Prop_OG_NOMA}, we next

In this section, we  present a power control strategy to speed up the convergence in Air-FedAvg by minimizing the optimality gap
in Proposition~\ref{Prop_OG_NOMA} with given numbers of outer iterations $T$ and local training epochs $\Omega$, via jointly optimizing the transmission power scaling factors $\{p_{k,t}\}$ at edge devices and denoising factors $\{\eta_t\}
$ at edge server for AirComp.
Mathematically, the optimality gap minimization problem is formulated as
 \begin{align}
	 	\mathbf{(P1):} ~\min_{\{p_{k,t}\ge 0\}, \{\eta_t\geq 0\}} ~~& {\Phi}\left(\{p_{k,t}\},\{\eta_t\} ,T,\Omega\right) \notag\\
 {\rm s.t.}~~~~~~& p_{k,t}  \leq  \tilde{P}^{\rm max}_k,~\forall k\in{\mathcal K}, ~ t\in\mathcal{T}
 \label{P1_P_max1}\\
& \frac{1}{T} \sum_{t=1}^{T}p_{k,t} \leq \tilde{P}^{\rm ave}_k,~\forall k\in{\mathcal K}\label{P1_P_ave1},
\end{align}
where $ \tilde{P}^{\rm max}_k=\frac{q P^{\rm max}_k}{W_k^2}$, $ \tilde{P}^{\rm ave}_k=\frac{q P^{\rm ave}_k}{W_k^2 }$, and the power constraints in \eqref{P1_P_max1} and \eqref{P1_P_ave1} directly follow  from \eqref{sys_bar_P_max1} and \eqref{sys_bar_P_ave1}, respectively.

By discarding the constant terms (i.e., the initial optimality gap $F\left({\bf v}_{0}\right)-F^{\star} $, the gradient variance term $V$, and the model divergence term $B$) in ${\Phi}\left(\{p_{k,t}\},\{\eta_t\} ,T,\Omega\right) $ for problem (P1), we have a recasted optimality gap minimization problem as
\begin{align}
	 	\mathbf{(P1.1):} ~~~~~~~~&\notag\\
	 	\min_{\{p_{k,t}\ge 0\}, \{\eta_t\geq 0\}} ~~& \sum_{t=1}^{T}\! \left(\!a_t  \!\!\sum\limits_{k\in\mathcal K}\! c_k\! \left(\frac{h_{k,t}\sqrt{p_{k,t}}}{\sqrt{\eta_t}}-1\right)^2+b_t\frac{\sigma_z^2q}{\eta_t}\right)\notag\\
 {\rm s.t.}~~~~~~& \eqref{P1_P_max1}~\text{and}~\eqref{P1_P_ave1},\notag
\end{align}
where $a_t=\frac{J_t}{2\gamma_{t-1}} +\frac{J_t(L+\gamma_{t-1}L^2\Omega)}{2}$, $b_t=\frac{J_t(L+\gamma_{t-1}L^2\Omega)}{2K^2}$, and $ c_k= \frac{W_k^2}{ K}, \forall k\in\mathcal{K}$.
 Note that problem (P1.1) or (P1) is highly non-convex due to the coupling of the transmission power scaling factors and the denoising factors in the objective function.
To deal with this difficulty, we adopt the alternating optimization technique to problem (P1.1), in which the denoising factor $\{\eta_t\}$ and the transmission power scaling factor $\{p_{k,t}\} $ are optimized iteratively in an alternating manner, by considering the other to be given in each iteration.

First, we optimize $\{\eta_t\}$ in problem (P1.1) under given $\{p_{k,t}\} $. In this case, problem (P1.1) is decomposed into the following $T$ subproblems each for one specific outer iteration $t\in\mathcal{T}$.
\begin{align}
	 	\min_{\eta_t\geq 0} ~~& a_t  \sum\limits_{k\in\mathcal K} c_k \left(\frac{h_{k,t}\sqrt{p_{k,t}}}{\sqrt{\eta_t}}-1\right)^2+b_t\frac{\sigma_z^2q}{\eta_t}.\label{NOMA_eta}
\end{align}
By introducing  $\hat{\eta}_t=1/\sqrt{\eta_t}$, problem \eqref{NOMA_eta} is recast as the following convex quadratic problem:
\begin{align}
	 	\min_{\hat{\eta}_t\geq 0} ~~& a_t  \sum\limits_{k\in\mathcal K} c_k \left(h_{k,t}\sqrt{p_{k,t}}\hat{\eta}_t-1\right)^2+b_t\sigma_z^2q \hat{\eta}^2_t\label{NOMA_hat_eta}.
\end{align}
By setting the first derivative of the objective function in problem \eqref{NOMA_hat_eta} to be zero, we can obtain the optimal solution $\hat{\eta}_t^*$ to problem \eqref{NOMA_hat_eta}, and accordingly get the optimal solution to problem \eqref{NOMA_eta} as ${\eta}_t^{\star}=(\frac{1}{\hat{\eta}_t^*})^2,\forall t\in\cal T$, given in the following proposition.
\begin{proposition}\label{Lemma_NOMA}\emph{
	With any given $\{p_{k,t}\} $, the optimal solution of ${\eta}_t $ to problem \eqref{NOMA_eta} is given by
\begin{align}\label{NOMA_eta_opt}
	{\eta}_t^{\star} =\left( \frac{a_t  \sum\limits_{k\in\mathcal K} c_k (h_{k,t})^2p_{k,t}+b_t\sigma_z^2q }{ a_t  \sum\limits_{k\in\mathcal K} c_k h_{k,t}\sqrt{p_{k,t}}} \right)^2,~t\in\mathcal{T}.
\end{align}
}
\end{proposition}

%, and then the objective $\tilde{\Phi}\left(\{p_{k,t}\},\{\eta_t\} \right)$ in problem (P1) under any given $\{\eta_t\}$ is reduced into
%\begin{align}
%	\hat{\Phi}\left(\{p_{k,t}\}\right)\triangleq  \sum_{i=1}^{T} a_t  \sum\limits_{k\in\mathcal K} c_k \left(\frac{h_{k,t}\sqrt{p_{k,t}}}{\sqrt{\eta_t}}-1\right)^2.
%\end{align}
%

%\subsubsection{transmission power scaling Factor Optimization under Given Denoising Factor}
% optimize $\{{p}_{k,t}\}$ under given
Next, we optimize the transmission power scaling variables $\{p_{k,t}\}$ in problem (P1.1) under given denoising factors $\{{\eta}_t\}$. In this case, problem (P1.1) reduces to
%Unfortunately, the resultant problem is a non-convex optimization problem in its original form:
\begin{align}
	 ~\min_{\{p_{k,t}\ge 0\}} ~~&  \sum_{i=1}^{T} a_t  \sum\limits_{k\in\mathcal K} c_k \left(\frac{h_{k,t}\sqrt{p_{k,t}}}{\sqrt{\eta_t}}-1\right)^2\label{NOMA_power}\\
 {\rm s.t.}~~~&\eqref{P1_P_max1}~\text{and} ~\eqref{P1_P_ave1}.\notag
\end{align}
Although problem  \eqref{NOMA_power} is non-convex in its current form, via a change of variables $ \hat{p}_{k,t}\triangleq\sqrt{p_{k,t}}, \forall k\in\mathcal{K}, t\in\mathcal{T}$, problem \eqref{NOMA_power} is equivalently transformed into the following convex form:
\begin{align}
 \min_{\{\hat{p}_{k,t}\ge 0\}} ~~&\sum_{i=1}^{T} a_t  \sum\limits_{k\in\mathcal K} c_k \left(\frac{h_{k,t}\hat{p}_{k,t}}{\sqrt{\eta_t}}-1\right)^2\label{NOMA_power_q}\\
 {\rm s.t.}~~~ &\hat{p}_{k,t} \leq \sqrt{ \tilde{P}^{\rm max}_k},~\forall k\in{\mathcal K}, ~ t\in\mathcal{T}\label{P1_q_max1}\\
& \frac{1}{T} \sum_{t=1}^{T} \hat{p}_{k,t}^2  \leq \tilde{P}^{\rm ave}_k,~\forall k\in{\mathcal K}\label{P1_q_ave1},
\end{align}
where constraints \eqref{P1_q_max1} and \eqref{P1_q_ave1} follow from \eqref{P1_P_max1} and \eqref{P1_P_ave1}, respectively.
%Notice that problem (P3.1) is convex and thus can be optimally solved.

%\subsubsection{Optimal Solution to Problem \eqref{NOMA_power_q}}
% Let $\{\hat{p}_{k,t}^{\star}\}$ denote the optimal solution to problem \eqref{NOMA_power_q}.
By leveraging the Lagrange duality method, we have the following lemma.
%where $\lambda_k^{\rm opt}$ is the optimal dual variable associated with the $k$-th constraint in \eqref{P1_P_ave1}.

\begin{lemma}\label{lemma_NOMA_power_q}\emph{The optimal solution to problem \eqref{NOMA_power_q} is given as
\begin{align}
\hat{p}_{k,t}^{\star}=\min \left[\frac{h_{k,t}\sqrt{\eta_t}}{h_{k,t}^2 + \frac{\eta_t\lambda_k^{\rm opt} }{a_t c_k T } },\sqrt{ \tilde{P}^{\rm max}_k}~\right], \forall k\in\mathcal{K},~t\in\mathcal{T},\label{Air_q_Opt}
\end{align}
where $\lambda_k^{\rm opt}$ is the optimal dual variable associated with the $k$-th constraint in \eqref{P1_q_ave1}, which can be obtained via solving the dual problem of problem \eqref{NOMA_power_q}.
}
\end{lemma}
\begin{IEEEproof}
See Appendix~\ref{Proof_lemma_NOMA_power_q}.
\end{IEEEproof}
From \eqref{Air_q_Opt} in Lemma~\ref{lemma_NOMA_power_q}, we obtain the optimal transmission power scaling variables $\{p_{k,t}^{\star}\}$ to problem \eqref{NOMA_power} as
\begin{align}\label{Air_power_Opt}
{p}_{k,t}^{\star}&=\left(\hat{p}_{k,t}^{\star}\right)^2\notag\\
&\!=\!\min\!\left[\!\left(\!\frac{h_{k,t}\sqrt{\eta_t}}{ h_{k,t}^2 \!+ \!\frac{\eta_t\lambda_k^{\rm opt}}{a_t c_k T } } \right)^2\!\!,{ \tilde{P}^{\rm max}_k}\!\right],\! \forall k\in\mathcal{K},~t\in\mathcal{T}.
\end{align}
\begin{remark}\label{Remark_q}\emph{It is observed from \eqref{Air_q_Opt} and \eqref{Air_power_Opt} that the optimal transmission power scaling variables $\{p_{k,t}^{\star}\}$ exhibit a \emph{regularized channel inversion} structure with the regularized term $\frac{\eta_t\lambda_k^{\rm opt}}{a_t c_k T }$, which is related to its average power budget in \eqref{P1_q_ave1} through the optimal dual variable $\lambda_k^{\rm opt}$. In particular, based on the complementary slackness condition for problem \eqref{NOMA_power_q}, it follows that
if $\lambda_k^{\rm opt}>0$ holds for edge device $k\in\mathcal{K}$, then we have $ \frac{1}{T} \sum_{t=1}^{T}{p}^{\star}_{k,t} - \tilde{P}^{\rm ave}_k=0 $, such that  this edge device should run out of its power with the regularized channel-inversion power control; otherwise, if $\lambda_k^{\rm opt}=0$, then edge device $k$ should transmit with channel-inversion transmission power control, without using up its power.
}
\end{remark}

Now, with the obtained $\{{\eta}_t^{\star}\}$ in \eqref{NOMA_eta_opt} and $ \{{p}_{k,t}^{\star}\}$ in \eqref{Air_power_Opt}, we present the complete algorithm to solve problem (P1), in which $ \{{p}_{k,t}\}$ and $\{{\eta}_t\}$ are updated alternately in an iterative manner, as shown in Algorithm 1.
In each iteration, we first solve problem \eqref{NOMA_eta} under given $\{ p_{k,t}\}$ to update $\{{\eta}_t\}$ as $\{{\eta}_t^{\star}\}$, and then solve \eqref{NOMA_power} under  $\{ \eta_{t}\}$ to update $\{ p_{k,t}\}$ as $ \{{p}_{k,t}^{\star}\}$.
It is easy to show that Algorithm 1 would lead to monotonically non-increasing objective values for (P1.1) over iterations. As the optimal value of problem (P1) is upper-bounded, it is evident that the convergence of Algorithm 1 can be ensured.
%Since the optimal value of problem (P1) is monotonically nondecreasing at each iteration. This together with the fact that the optimal value of problem (P1) is upper-bounded suggests that the alternating-optimization-based algorithm will converge to at least a locally optimal solution to problem (P1).
%In summary, the proposed iterative algorithm is presented in Algorithm 2.

\begin{table}[htp]
%\vspace{0.1cm} \textbf{Algorithm 1}  \vspace{0.1cm}
\begin{center}%\vspace{-0.1cm}
\hrule
\normalsize
\vspace{0.2cm} \textbf{Algorithm 1 for Solving Problem (P1.1)}\vspace{0.2cm}
\hrule \vspace{0.1cm} % \hrule
\begin{itemize}
    \item[1]  Initialization: Set the initial power control $\{p_{k,t}^{(0)}\}$ and $i=0$.
    \item[2]  {\bf Repeat:}
                \begin{itemize}
                \item[a)]  With given $ p_{k,t}=p_{k,t}^{(i)}, \forall k\in\mathcal{K}, t\in\mathcal{T}$, obtain the optimal solution to problem \eqref{NOMA_eta} as $\eta_t^{(i)}={\eta}_t^{\star},~\forall t\in\mathcal{T}$ in \eqref{NOMA_eta_opt};
                 \item[b)]  With given $ \{{\eta}_t^{(i)}\}$, obtain the optimal solution to  problem \eqref{NOMA_power} as $p_{k,t}^{(i)}={p}_{k,t}^{\star}, \forall k\in\mathcal{K}, t\in\mathcal{T}$ in \eqref{Air_power_Opt};
                \item[c)] Set $p_{k,t}^{(i+1)}={p}_{k,t}^{\star}, \forall k\in\mathcal{K}, t\in\mathcal{T}$, and $i=i+1$.
                \end{itemize}
     \item[3] {\bf Until} the objective value of problem (P1.1) converges within a given threshold.
    \end{itemize}
%\vspace{0.1cm}
\hrule \vspace{0cm}
\end{center}\vspace{-1cm}
\end{table}

\section{Training Latency Minimization: Air-FedAvg versus OMA-FedAvg}

Besides the optimality gap or training accuracy, the training latency is another important metric to measure the learning performance.
This section investigates the training latency performance of Air-FedAvg while ensuring a given optimality gap requirement, and compares it with the conventional OMA-FedAvg \cite{MChen21_TWC,Mo2020aa,Tran19_Infocom} to show the benefit of the AirComp-enabled NOMA over OMA for FEEL.
 For fair comparison on the training latency, we consider both  the communication latency for model update aggregation and the computation latency for local training updates.
In the following two subsections, we minimize the training latency for Air-FedAvg and OMA-FedAvg, respectively, by jointly optimizing the power control $\{p_{k,t}\}$ and $\{\eta_t\}$, together with the numbers of outer iterations $T$ and local training epochs $\Omega$.

% we aim to focus on the training latency  minimization with a common target on the optimality gap derived in  Theorem~\ref{Theo_OG_NOMA} for both Air-FedAvg and OMA-FedAvg schemes as performance comparison, respectively.
%Note that the training latency includes the transmission and computation latency, and thus the total number of global updates $T$ and local updates at each round $\Omega$ should be taking into account.

\subsection{Training Latency Minimization for Air-FedAvg} \label{Air-FedAvg}

First, we consider the case with Air-FedAvg. Recall that the size of model parameters is denoted by $q$ and assume that each element of model parameter is modulated as a single analog symbol.
%, and denote the bandwidth for AirComp as $B$.
 Therefore, to upload a model update, the total number of analog symbols to be transmitted is $q$.
Let $M$ represent the number of symbols in each resource block with duration $T_{\rm slot}$.
Thus, for each outer iteration, the transmission latency is thus expressed as
\begin{align}\label{Latency_Air}
    \tau_{\rm tran}=\text{ceil}\left(\frac{q}{M}\right)T_{\rm slot},
\end{align}
where $\text{ceil}(\cdot)$ is the integer ceiling function.
For instance, in LTE systems \cite{LTE}, each resource block with duration of $T_{\rm slot}=1$ ms  consists two slots with $14$ symbols in general, and thus we have  $M=14$ in LTE systems.

%Hence, the per-iteration latency for Air-FedAvg is given by
%\begin{align}\label{Latency_Air}
%    \tau_{\rm tran}=\frac{q}{M}T_{\rm slot}.
%\end{align}
%For each outer iteration, as each parameter element is transmitted over one symbol, the transmission latency is thus expressed as $\tau_{\rm tran} = q/B$.

%Let $\rho $ denote the target optimality gap, where the optimality gap is given by  $ \Phi\left(\{p_{k,t}\},\{\eta_t\},T, \Omega \right)$ in \eqref{Power_OG_NOMA}.
% Assume $\tau_{\rm tran}$ as the transmission time at each round for concurrently model uploading from all edge devices to the edge server via AirComp scheme.
 Furthermore, the computation latency at each outer iteration corresponds to that required for the $K$ devices to implement $\Omega$ local training epochs. Let $c$ denote the {\it central processing unit} (CPU) cycles for computing each sample and $f$ denote the CPU frequency.
 As there are a total of $|\tilde{\mathcal D}_k|$ samples to be processed, the computation latency is thus expressed as $\tau_{\rm comp} = {c |\tilde{\mathcal{D}}_k|}/{f}$.
 In this case, the total training latency for Air-FedAvg is $\Delta^{\rm A}=T\left( \tau_{\rm tran}+ \tau_{\rm comp}\Omega\right)$\footnote{We ignore the computation time at the edge server and the transmission/broadcasting time from the edge server to the $K$ edge devices, as the edge server normally has strong computation power and high transmission power, thus making its computation and broadcasting time negligible constants that can be omitted. }.

 Then, the training latency minimization problem under a targeted optimality gap $\rho$  is formulated as (P2) as follows, by jointly optimizing the transmission power scaling factors $\{p_{k,t}\}$, denoising factors $\{\eta_t\}$, and the numbers of outer iterations $T$ and local training epochs $\Omega$.
\begin{align}
	 	\mathbf{(P2):} \min_{\substack{\{p_{k,t}\ge 0\}, \{\eta_t\geq 0\}, \\T\in\mathbb{Z}^{+}, \Omega\in\mathbb{Z}^{+} }} & T\left( \tau_{\rm tran}+\tau_{\rm comp}\Omega\right) \notag\\
 {\rm s.t.}~~~~~~& \Phi\left(\{p_{k,t}\},\{\eta_t\},T ,\Omega\right)\leq \rho \label{P1_Target_OG}\\
& \eqref{P1_P_max1}~\text{and}~\eqref{P1_P_ave1},\notag
\end{align}
where $\mathbb{Z}^{+}$ denotes the set of all positive integers.
Notice that problem (P2) is a mixed integer nonlinear program, which is challenging to solve. To deal with this issue, we show that problem (P2) can be equivalently solved via solving the following feasibility checking problem (P2.1) under given $T$.
\begin{align}
	 	\mathbf{(P2.1):} ~~\text{find}&~~\Omega, ~ \{p_{k,t}\ge 0\},~\text{and}~\{\eta_t\geq 0\} \notag\\
 {\rm s.t.}&~~~ \eqref{P1_P_max1}, ~\eqref{P1_P_ave1},~\text{and}~\eqref{P1_Target_OG}.\notag
\end{align}
Suppose that the optimal solution of $T$ to problem (P2) is $T^{\rm A,\star}$. Then, under any given $T$, if problem (P2.1) is feasible, then it follows that  $T^{\rm A, \star}\leq T$; otherwise $T^{\rm A, \star}\geq T$. Therefore, we can solve problem (P2) by first checking the feasibility of problem (P2.1) under any given $T$ and then using a bisection search over $T$.
Therefore, in the following, we only need to consider the feasibility checking problem (P2.1) under given $T$.

%we first optimize $\{p_{k,t}\}$ and $\{\eta_t\}$, and $\Omega$ under given $T$, and then search over $T$.

%non-convex optimization problem, as the optimization variable $T$ is an integer, and constraint \eqref{P1_Target_OG} is non-convex due to the coupling of the  transmission power scaling and denoising factors. Furthermore, since $T$ is a-priori unknown, problem (P2) consists of an uncertain number of constraints \eqref{P1_P_max1}.
%Therefore, problem (P2) is difficult to be solved directly.

%\subsubsection{Optimization}
%To overcome this challenge, we apply an iterative algorithm to solve problem (P2) by first jointly optimizing the the transmission power scaling factors $\{p_{k,t}\}$ at edge devices and denoising factor $ \{\eta_t\}$ at edge server and then using a bisection search to find the optimal number of outer iterations denoted by  $T^{\rm A, \star}$.

Now, consider problem (P2.1), which can be equivalently solved via solving the optimality gap minimization problem as follows:
%Let $\hat{\rho}^{\star}$ denote the optimal value of problem (P2.2). Then, solving problem (P2.2) is equivalent to solving the following optimality gap minimization problem under given $T$:
\begin{align}
	 	\mathbf{(P2.2):} ~\hat{\rho}^{\star}=\min_{\{p_{k,t}\ge 0\}, \{\eta_t\geq 0\},\Omega\in\mathbb{Z}^{+}} ~~& \Phi\left(\{p_{k,t}\},\{\eta_t\},T,\Omega \right) \notag\\
 {\rm s.t.}~~~~~~~~~~&\eqref{P1_P_max1}~\text{and} ~\eqref{P1_P_ave1}.\notag
\end{align}
Notice that if $\hat{\rho}^{\star}\leq {\rho}$, then problem (P2.1) is feasible; and otherwise infeasible.
Notice that with any given $ \Omega$, problem (P2.2) reduces to problem (P1), which can be solved via Algorithm 1.
Therefore, we can solve problem (P2.2) by first finding the optimized $\{p_{k,t}\}$ and $\{\eta_t\}$ under any given $\Omega$ via Algorithm 1, and then adopting a one-dimension search to find a desired $\Omega$ achieving the minimum objective value.  Notice that the alternating-optimization-based Algorithm 1 only achieves a converged but generally sub-optimal solution to problem (P1), and as a result,  the proposed solution to problem (P2) is sub-optimal in general as well, though with high quality as will be shown in simulations next.

%Thus, with problem (P1) solved and via a one-dimension search over $\Omega$, the feasibility problem (P2.1) (equivalent to problem (P2.1)) is accordingly checked.
%By combing this together with the bisection search over $T$ to find $T^{\rm A, \star}$, problem (P2) can be efficiently solved.
%Notice that due to the fact that under given $T$, we only obtain a locally optimal solution to problem (P1), the obtained solution to problem (P2) is generally suboptimal.

\subsection{Training Latency Minimization for OMA-FedAvg}\label{OMA_Fed}
Next, for comparison, we consider the OMA-FedAvg, in which the TDMA protocol is adopted for multiple edge devices to upload their local models digitally over orthogonal time slots\footnote{
Notice that under our considered single-antenna setup, the performance gain by the digital power-domain NOMA \cite{YLiu17_NOMA,ZDing17_Jsac,Liu2021_NGMA} over the OMA scheme (e.g., the considered TDMA) might become marginal (e.g., in terms of spatial multiplexing gain). In this case and for initial investigation, we only consider the OMA scheme as the benchmarking scheme to show the superiority of the AirComp in terms of aggregation latency in the context of FEEL. The comparison with digital NOMA is left for future work. }.

In particular, to facilitate the digital uploading, each edge device $k\in \mathcal K $ employs a random quantizer (or a stochastic low-precision quantizer) on the updated local model $\tilde{\bf w}_{k,t} \in \mathbb{R}^q$, which is denoted as  $\mathcal{Q}( {\bf x}): \mathbb{R}^q\rightarrow\mathbb{R}^q$ with \cite{Quantizer_reisizadeh20a,NIPS2017_QSGD}
\begin{align}
	\mathcal{Q}( {x_i})=\| {\bf x}\| \cdot {\rm sign} \left( x_i\right)\cdot Q_i\left({\bf x},s \right), ~\forall i\in\{1,\cdots,q\}.
\end{align}
Here, $Q_i\left({\bf x},s \right)$ denotes a random variable defined as
\begin{align}
	Q_i\left({\bf x},s \right)=	
	\begin{cases}
 (l+1)/{s},~&{\rm with~probability~} \frac{|x_i|}{\|{\bf x} \|}s-l,\\
	l/{s},~&{\rm otherwise},
	\end{cases}\label{Quantize_Q}
\end{align}
where  $s$ is a tuning parameter defining the number of quantization levels and $0\leq l\leq s$ is an integer such that $\frac{|x_i|}{\|{\bf x} \|}\in\left[\frac{l}{s},\frac{l+1}{s} \right] $. Let $\hat{q}=\min\left\{ \frac{\sqrt{q}}{s} ,\frac{q}{s^2}\right\}$, and it has been shown in \cite{Quantizer_reisizadeh20a,NIPS2017_QSGD} that $\mathcal{Q}( {\bf x})$ is unbiased and its MSE is bounded, i.e.,
\begin{align}
	&\mathbb{E}\left(\mathcal{Q}( {\bf x}) \right)={\bf x},\label{Quantize_Bias}\\
	&\mathbb{E}\left(\left\|\mathcal{Q}( {\bf x})-{\bf x} \right\|^2\right)\leq \hat{q}\left\|{\bf x} \right\|^2.\label{Quantize_MSE}
\end{align}
%where $\hat{q}=\min\left\{ \frac{\sqrt{q}}{s} ,\frac{q}{s^2}\right\}$.
%}	
%\end{definition}

Based on the quantizer, the transmitted signal by device $k$ at each outer iteration $t$ is $\mathcal{Q}\left(\tilde{\bf w}_{k,t}\right)$, which includes three components, including the vector norm $\| \tilde{\bf w}_{k,t} \|$, the sign of each entry ${\rm sign} \left( \tilde{w}_{k,t,i}\right) $ with $\tilde{w}_{k,t,i}$ denoting the $i$-element of $\tilde{\bf w}_{k,t}$, and the quantization value of each entry $Q_i\left(\tilde{\bf w}_{k,t},s \right)$.
Suppose that $S_0$ bits are needed to encode  $\| \tilde{\bf w}_{k,t}\|$, $q$ bits for encoding $\{{\rm sign} \left( \tilde{w}_{k,t,i}\right) \}$,  and $q\log_2 s$ bits for $\{Q_i\left(\tilde{\bf w}_{k,t},s \right)\}$. Therefore, the totally required number of bits for edge device $k\in \mathcal K$ to send $\mathcal{Q}\left(\tilde{\bf w}_{k,t}\right)$ to the edge server is
\begin{align}
	S=(1+\log_2 s)q+S_0 \label{OMA_Bits}.
\end{align}
Then the achievable rate of device $k$ at outer iteration $t$ is given as
\begin{align}\label{OMA_rate}
	r_{k,t}(p_{k,t} )=B\log \left(1+\frac{p_{k,t}|h_{k,t}|^2}{\sigma_z^2} \right),
\end{align}
where $B$ denotes the bandwidth.
%Accordingly, the total aggregation delay in OMA-FedAvg is given by $\sum\limits_{t\in\mathcal{T}} \left(\sum\limits_{k\in\mathcal{K}}\tau_{k,t}+\tau_{\rm comp} \Omega \right)$.
 Let $\tau_{k,t}$ denote the transmission time for edge device $k$ uploading its model to the edge server via TDMA  at iteration $t$.
 In order for each edge device $k$ successfully uploading $S$ bits to edge server, it requires that
 \begin{align}\label{sys_rate}
 	\tau_{k,t}r_{k,t}(p_{k,t} )\geq S, \forall k\in\mathcal{K}, t\in\mathcal{T}.
 \end{align}
Besides the communication for model uploading, the computation latency for local training epochs is same as $\tau_{\rm comp}\Omega$ in Air-FedAvg. By combining them, the total training latency for OMA-FedAvg is
\begin{align}
	\Delta^{\rm O}=\sum\limits_{t\in\mathcal{T}} \left(\sum\limits_{k\in\mathcal{K}}\tau_{k,t}+\tau_{\rm comp} \Omega \right).
\end{align}

%Besides the communication for model uploading, the computation latency for local SGD iterations is same as $\tau_{\rm comp}\Omega$ in Air-FedAvg. By combining them, the total training latency for OMA-FedAvg is $\sum\limits_{t\in\mathcal{T}} \left(\sum\limits_{k\in\mathcal{K}}\tau_{k,t}+\tau_{\rm comp} \Omega \right)$.

Next, we consider the optimality gap achieved by OMA-FedAvg, in which certain quantization errors are introduced in \eqref{Quantize_Q}.
By replacing $\left\|\mathbb{E}\left({\bm \varepsilon}_{t}\right)\right\|^2$ and $\mathbb{E}\left(\left\|{\bm \varepsilon}_{t}\right\|^2\right)$ in Theorem \ref{Theo_OG_NOMA} as $\mathbb{E}\left({\bm \varepsilon}_{t}\right)=\frac{1}{K}\sum\limits_{k\in\mathcal{K}}\mathbb{E}\left(\mathcal{Q}( { \tilde{\bf w}_{k,t}})- { \tilde{\bf w}_{k,t}}\right) =0$ and $\mathbb{E}\left(\left\|{\bm \varepsilon}_{t}\right\|^2\right)\leq\frac{1}{K}\sum\limits_{k\in\mathcal{K}}\mathbb{E}\left(\left\|\mathcal{Q}( { \tilde{\bf w}_{k,t}})- { \tilde{\bf w}_{k,t}}\right\|\right) \leq \frac{\hat{q}}{K}\sum\limits_{k\in\mathcal{K}} W_k^2$ in \eqref{Quantize_Bias} and \eqref{Quantize_MSE}, respectively, the expectation of optimality gap satisfies that
\begin{align}\label{OG_OMA}
	&\mathbb{E}\left(F\left({\bf v}_{T}\right)\right)-F^{\star}\leq	\Theta\left(T, \Omega \right)\triangleq
\notag\\
	& ~~~\prod_{t\in\mathcal{T}}C_t \left(F\left({\bf v}_{0}\right)-F^{\star}\right)  +\sum_{t=1}^{T} J_t \left( \gamma_{t-1} B+  \gamma_{t-1}^2 V +\right)\notag\\
	&~~~~+\sum_{t=1}^{T} J_t \left( \frac{\left(L+\gamma_{t-1}L^2\Omega\right)\hat{q}}{2K}\sum\limits_{k\in\mathcal K} W_k^2\right).
\end{align}

 \begin{remark}\emph{
	From \eqref{OMA_Bits}, \eqref{sys_rate}, and \eqref{OG_OMA}, there exists a fundamental trade-off in designing the quantization level $s$. In particular, if $s$ is small, then the transmission time $\tau_{k,t}$'s would reduce but the corresponding quantization errors would be amplified, thus leading to the compromised optimality gap.}
\end{remark}

Now, with the training latency and the optimality gap upper bound in \eqref{OG_OMA}, the optimality gap constrained training latency minimization problem for OMA-FedAvg is formulated as follows, in which the numbers of outer iteration $T$ and local training epochs $\Omega$, transmission power control $\{p_{k,t}\}$, and time allocation $ \{\tau_{k,t}\}$ are jointly optimized:
\begin{align}
	 	\mathbf{(P3):}~~~~~~~~&\notag\\
	 	 \min_{\substack{\{p_{k,t}\ge 0\}, \{\tau_{k,t}\geq 0\}, \\T\in\mathbb{Z}^{+}, \Omega\in\mathbb{Z}^{+} }} ~~& \sum\limits_{t\in\mathcal{T}} \left(\sum\limits_{k\in\mathcal{K}}\tau_{k,t}+\tau_{\rm comp} \Omega \right) \notag\\
 {\rm s.t.}~~~~~~~~~& p_{k,t}  \leq  \tilde{P}^{\rm max,O}_k,~\forall k\in{\mathcal K}, ~ t\in\mathcal{T}\label{P3_P_max1}\\
& \frac{1}{T} \sum_{t=1}^{T}p_{k,t} \leq \tilde{P}^{\rm ave,O}_k,~\forall k\in{\mathcal K}\label{P3_P_ave1}\\
& 	\tau_{k,t}r_{k,t}(p_{k,t})\geq S, \forall k\in\mathcal{K}, t\in\mathcal{T}\label{P3_rate}\\
& \Theta\left(T,\Omega \right)\leq \rho, \label{P3_Target_OG}
\end{align}
where \eqref{P3_P_max1} and \eqref{P3_P_ave1} denote individual maximum and average transmission power constraints, respectively, \eqref{P3_rate} denotes the rate constraints for model uploading, and constraint \eqref{P3_Target_OG} ensures the maximum optimality gap requirement.
Similar to problem (P1), problem (P3) is a mixed integer non-linear program that is difficult to solve.

Next, we solve problem (P3) by first optimizing $\{p_{k,t}\}$ and $\{\tau_{k,t}\}$ under any given feasible $T$ and $\Omega$, and then searching over $T$ and $\Omega$. With given $T$ and $\Omega$, problem (P3.1) reduces as
\begin{align}
	 	&\mathbf{(P3.1):} ~~~\notag\\
	 	&\min_{\substack{\{p_{k,t}\ge 0\}, \\ \{\tau_{k,t}\geq 0\}}} ~ \sum_{t=1}^{ T} \sum\limits_{k\in\mathcal{K}} \tau_{k,t} \notag\\
 &~~~~{\rm s.t.}~~~ \!B\log \!\left(\!1\!+\!\frac{p_{k,t}\left|h_{k,t}\right|^2}{\sigma_k^2q} \!\right)\!\geq \frac{S}{	\tau_{k,t}}\!, \forall k\in\mathcal{K}, t\in\mathcal{T}\label{P3_rate2}\\
 & ~~~~~~~~~~\eqref{P3_P_max1}~\text{and}~ \eqref{P3_P_ave1}, \notag
\end{align}
which is convex and thus can be directly solved optimally via standard convex optimization tools such as CVX \cite{cvx}.

%Let $\Gamma(T,\Omega)$ denote the obtained training latency under given $T$ and $\Omega$.

Then, we search over $T$ and $\Omega$. It is observed from problem (P3) that under any given $\Omega$, the training latency is monotonically increasing w.r.t. $T$ (provided that $T$ is feasible or satisfies \eqref{P3_Target_OG}).
Therefore, we can find the desirable $T$ and $\Omega$ by first finding the minimum $T$  satisfying \eqref{P3_Target_OG} under any given $\Omega$, and then using a one-dimensional search to find the optimal $\Omega$ achieving the minimum training latency (the optimal value of problem (P3.1) under given $T$ and $\Omega$). By implementing this, the optimal $T$ and $\Omega$ can be found, and the corresponding optimal solution of $\{p_{k,t}\}$ and $\{\tau_{k,t}\}$  becomes optimal for (P3). As a result, problem (P3) is finally solved.

\subsection{Training Latency Comparison Between Air-FedAvg and OMA-FedAvg}\label{AirVsOMA}
%{\color{blue}
With the minimally achievable training latencies obtained in Sections \ref{Air-FedAvg} and \ref{OMA_Fed}, this subsection provides a comparison between them to provide insights on the benefit of Air-FedAvg over OMA-FedAvg.
In the following, we compare their achieved the per-iteration communication latency and the number of the outer iterations for the two systems, respectively, which are critical for the total training latency.
%}

%\subsubsection{per-iteration Latency Comparison}
%For Air-FedAvg, we assume that each element of model parameter is modulated as a single analog symbol. To upload a model update with dimension $q$, the total number of analog symbols to be transmitted is calculated as $q$ symbols.
%Let $M$ represent the transmission symbol in each channel use. For instant, in LTE network, there exist at least $6$ resource blocks \cite{LTE}. Each resource block with duration of $T_{\rm slot}=1$ (ms)  is consisted of two subframes with $14$ symbols in general, and thus  $M=14\times6$ in LTE system. Hence, the per-iteration latency for Air-FedAvg is given by
%\begin{align}\label{Latency_Air}
%    \tau_{\rm tran}=\frac{q}{M}T_{\rm slot}.
%\end{align}
First, we consider the per-iteration communication latency. For Air-FedAvg, it is observed from \eqref{Latency_Air} that  the transmission latency at each outer iteration is proportional to the model size $q$, but independent of the number of edge devices $K$.
By contrast, in OMA-FedAvg, it is observed from \eqref{OMA_rate}  that the total transmission latency of the $K$ edge devices at each iteration is given by
\begin{align}\label{Latency_OMA}
	\tau_t^{\rm O}\triangleq \sum\limits_{k\in\mathcal{K}}\tau_{k,t}& = \sum\limits_{k\in\mathcal{K}} \frac{S}{r_{k,t}(p_{k,t})} \notag\\
	&= \sum\limits_{k\in\mathcal{K}} \frac{(1+\log_2 s)q+S_0}{r_{k,t}(p_{k,t})},
\end{align}
which scales approximately linearly with $K$ and critically depends on the value of quantization level $s$. It is evident that Air-FedAvg is beneficial in reducing the communication latency at each iteration, especially when $K$ becomes large.
%the total offloaded bits for each edge device to upload a model update with dimension $q$ after proper quantization operation is $S$ bits in \eqref{OMA_Bits}.
%Accordingly, the transmission time of edge device $k$ at outer iteration $t$ is $ \tau_{k,t}=\frac{S}{r_{k,t}}$, and the total transmission latency is given by
%\begin{align}\label{Latency_OMA}
%	\tau_t^{\rm O}\triangleq \sum\limits_{k\in\mathcal{K}}\tau_{k,t} = \sum\limits_{k\in\mathcal{K}} \frac{(1+\log_2 s)q+S_0}{r_{k,t}}  .
%\end{align}
%It is apparently that the latency of OMA-FedAvg scales approximately linearly with $K$.

%Combining the above two latency, the latency-reduction ratio of the OMA-FedAvg w.r.t. the Air-FedAvg is conducted as follows:
%\begin{align}\label{Lat_Ratio}
%	\frac{\tau_t^{\rm O}}{ \tau_{\rm tran}}\triangleq \frac{M\left(q+q\log_2 s+S_0\right)\sum\limits_{k\in\mathcal{K}} \frac{1}{r_{k,t}}}{q T_{\rm slot} }.
%\end{align}
%
%
%\begin{remark}\emph{
%	From \eqref{Lat_Ratio}, the gain of latency reduction for Air-FedAvg is more related to the quantization level $s$ and the number of edge devices $K$. However, there exists a fundamental trade-off in $s$. If $s$ is small, the transmission time of OMA-FedAvg would be reduce but its quantization error would be amplified. As for $K$, the transmission time for OMA-FedAvg is scaling linearly with $K$ increasing without any impact on per-iteration latency for Air-FedAvg. Thus the reduction gain is more prominent as $K$ grows.}
%\end{remark}

%{\color{blue}
Next, notice that the difference in number of outer iterations between Air-FedAvg and OMA-FedAvg arises from the distinct aggregation errors, i.e., the AirComp error for the former and the quantization error for the latter. Thus it is sufficient to compare the aggregation errors encountered by the two schemes. In general, the AirComp errors can be larger than the quantization errors when the channel quality is poor and the quantization level $s$ is large enough.
%Thus the needed number of outer iterations in Air-FedAvg is compariable  than that in OMA-FedAvg.
%otherwise, the opposite would hold due to the optimization of transmission power control.
%quantization errors would  the AirComp errors would  could be suppressed by .
However, at the extreme case when $K$ goes to sufficiently large and the power budgets are also large enough, it is observed from \eqref{equ_bias_p} and \eqref{equ_unbias_p} that the bias and MSE of aggregation errors would vanish with proper power control.
By contrast, the quantization error for the OMA-FedAvg scheme cannot vanish even in the large-$K$ regime.
 This shows the superiority of Air-FedAvg over OMA-FedAvg in reducing aggregation errors and the number of outer iterations, especially in large-scale networks with many participating edge devices.

\section{Simulation Results}\label{sec_simu}
%linear regression and handwritten digit identification
This section provides simulation results to validate the performance of the proposed power control design.
 In the simulation, the wireless channels from the edge devices to the edge server over different communication rounds follow {\it independent and identically distributed} (i.i.d.) Rayleigh fading, i.e., $\hat h_{k,t}$'s are modeled as i.i.d. {\it circularly symmetric complex Gaussian} (CSCG) random variables with zero mean and unit variance. We set the number of devices as $K=10$, the noise variance $\sigma_z^2=1$ W, and the average power budgets at different devices to be $\tilde{P}^{\rm ave}_k=1$ W, $\forall k\in\mathcal{K}$.
We set the maximum power budget to be $\tilde{P}^{\rm max}_k=5$ W and the diminishing learning rates $\eta_t=\frac{\beta}{t+a}$ with $a=10$ and $\beta=1$.

\begin{figure}[t]  \vspace{-0.05cm}
  \centering
  \subfigure[Optimality gap versus $T$ under $K=10$.]
  {\label{T_fig:OG}\includegraphics[width=8.8cm]{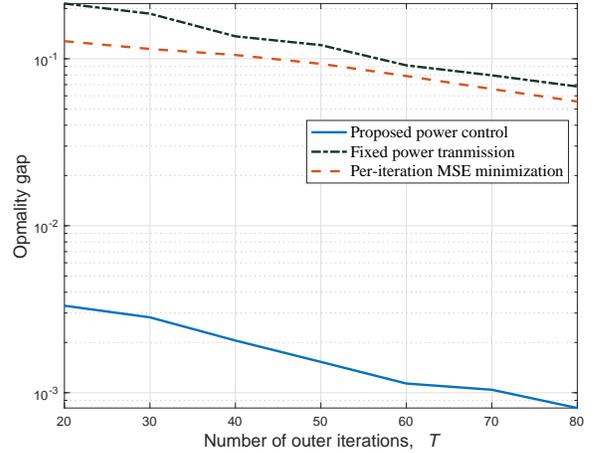}}
%  \vspace{0.08in}
  \subfigure[Prediction error versus $T$ under $K=10$.]
  {\label{T_fig:PE}
\includegraphics[width=8.8cm]{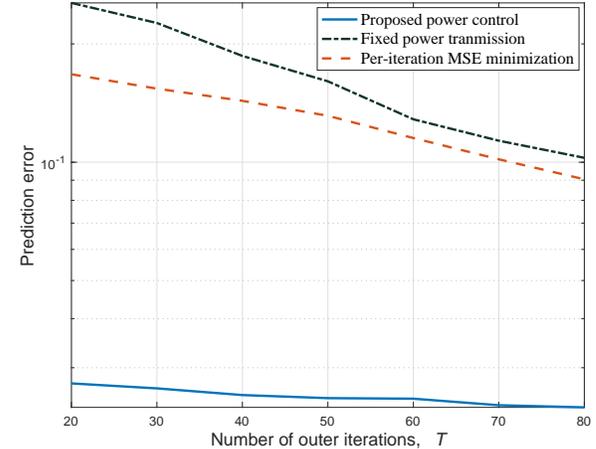}}
  \caption{Effect of the number of outer iterations on the learning performance of  Air-FedAvg with $\Omega^{\star}=5$.}
  \label{T_fig}
\vspace{-0.3cm}
\end{figure}
\begin{figure}[t]  %\vspace{-0.05cm}
  \centering
  \subfigure[Optimality gap versus $K$ under $T=50$.]
  {\label{K_fig:OG}\includegraphics[width=8.8cm]{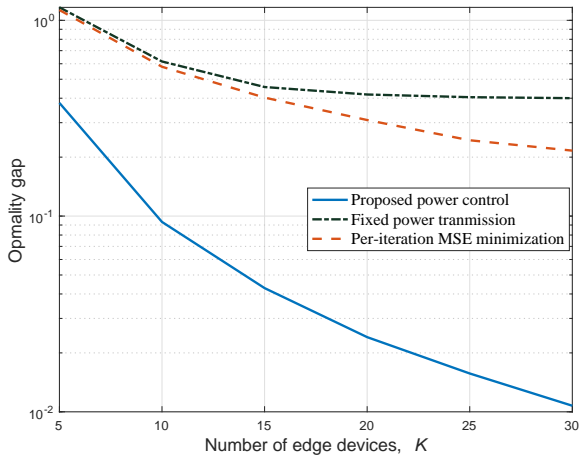}}
%  \vspace{0.08in}
  \subfigure[Prediction error versus $K$ under $T=50$.]
  {\label{K_fig:PE}
\includegraphics[width=8.8cm]{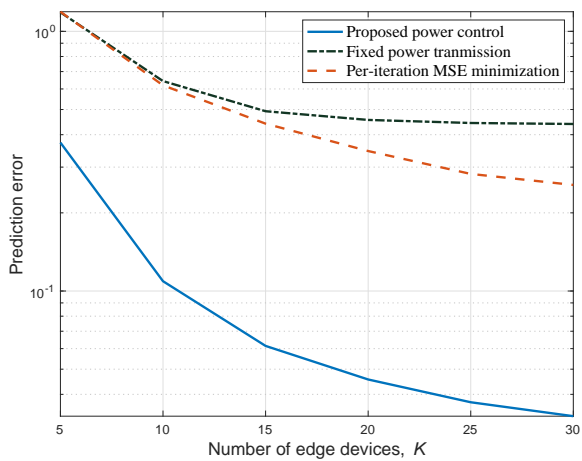}}
  \caption{Effect of the number of devices on the learning performance of Air-FedAvg with $\Omega^{\star}=5$.}
  \label{K_fig}
\vspace{-0.3cm}
\end{figure}

\subsection{Air-FedAvg for Ridge Regression and  Handwritten Digit Recognition}
First, we evaluate the performance of Air-FedAvg, for which the proposed power control policy is implemented using Matlab and Pytorch for ridge regression and handwritten digit recognition, respectively.
Furthermore, the optimality gap and prediction error are considered to measure the learning performance for ridge regression on synthetic dataset, while the loss function value and test (recognition) accuracy are considered for handwritten digit recognition on MNIST dataset.
%For performance comparison, we consider two benchmark schemes, .
We consider two benchmark schemes for performance comparison, namely the {\it fixed power transmission} that transmits with uniform power over different outer iterations by setting $p_{k,t}=\tilde{P}^{\rm ave}_k, \forall k\in\cal K$, and the {\it per-iteration MSE minimization} that minimizes the aggregation MSE independently at each outer iteration, which has been already solved in \cite{Cao_PowerTWC}.

%\subsection{Air-FedAvg for Ridge Regression}
First, consider the ridge regression, where the sample-wise loss function is $f({\bf w},{\bf x},y)=\frac{1}{2}\| {\bf x}^T{\bf w}-y\|^2$.
%with a regularization function $R({\bf w})=\|{\bf w}\|^2$ with the hyperparameter  $\rho=5\times 10^{-5}$.
Randomly generated synthetic dataset is used for model training and testing.
The generated data sample vector ${\bf x}\in\mathbb{R}^q$ with $q=20$ follows i.i.d. Gaussian distribution (i.e., $ x \sim\mathcal{N}(0,{\bf I})$) and the label $y$ is obtained as $y=x(2)+3x(5)+0.2z$, where $x(i)$ represents the $i$-th entry in vector ${\bf x}$ and $z$ denotes the observation noise with i.i.d. Gaussian distribution, i.e., $z\sim\mathcal{N}(0,1)$.
Unless stated otherwise, the data samples are evenly distributed among devices with identical size $ \bar{D} = 1000, \forall k\in \mathcal K$ and $| \mathcal{D}|=K\bar{D}$. The size of mini-batch is $n_b=500$.
Building on the above models, the smoothness parameter $L$ and Polyak-{\L}ojasiewicz parameter $\mu$ is obtained as the largest and smallest eigenvalues of the data Gramian matrix ${\bf X}^{\dagger}{\bf X}/| \mathcal{D}|+10^{-4}{\bf I}$, where ${\bf X}=[{\bf x}_1,\cdots,{\bf x}_{| \mathcal{D}|}]^{\dagger}$ is the data matrix.
The minimum value of loss function $F^{\star}$ is computed using the optimal parameter vector $\bf w^{\star}$ to the learning problem \eqref{OptimalParameter}, where ${\bf w}^{\star}=({\bf X}^{\dagger}{\bf X})^{-1}{\bf X}^{\dagger}{\bf y}$ with ${\bf y}=[y_1,\cdots,y_{| \mathcal{D}|}]^{\dagger}$.
We use the  simple upper bounds $G_k =2W_kL, \forall k\in\mathcal{K} $  \cite{DLiu2020Ar}, with $ W_k^2>\|{\bf w}^{\star} \|^2$, and the initial parameter vector is set as an all-zero vector.

Fig.~\ref{T_fig} shows the learning performance versus the number of outer iterations $T$  with  $K=10$, where the number of local training epochs is optimized as $\Omega^{\star}=5$.
First, it is observed that the proposed power control policy and the per-iteration MSE minimization design achieve faster convergence and lower optimality gap than the fixed power transmission design. This shows the benefit of power control optimization in accelerating the learning convergence rate, via directly minimizing the optimality gap.
Secondly, the proposed power control policy is observed to significantly outperform the per-iteration MSE minimization design in reducing the optimality gap. This is because that the contributions of model aggregation errors to the optimality gap are distinct over different outer iterations (see Remark~\ref{Remark_Theorem1}), which cannot be captured by the per-iteration MSE minimization design.
%Furthermore, the unbiased aggregation is observed to achieve a lower optimality gap than the biased aggregation when $N>150$ under the fixed learning rate and $N>80$ under the  diminishing learning rates. This coincides with Remark~\ref{Remark_ContractionRegion} that the FEEL algorithm converges to the optimal point with unbiased gradient aggregation.
%In addition, by comparing Fig. 3(a) (or  3(b)) versus Fig. 3(c) (or 3(d)), both the biased- and unbiased-aggregation schemes could achieve faster convergence in the diminishing learning rate case as depicted in Figs.~\ref{DL_fig:OG_v_N} and~\ref{DL_fig:PE_v_N}, than that in the fixed learning rate case as shown in Figs.~\ref{fig:OG_v_N} and~\ref{fig:PE_v_N}, which benefited from the adaptive learning rate.

Fig.~\ref{K_fig} shows the learning performance (i.e., the optimality gap in Fig.~\ref{K_fig:OG} and the prediction error in Fig.~\ref{K_fig:PE}) versus the number of devices $K$ with $T=50$ and $\Omega^{\star}=5$.
Firstly, it is observed that as $K$ increases, the optimality gap and prediction error achieved by all the three schemes decrease. This is due to the fact that the edge server can aggregate more data for averaging to improve the learning performance.
Secondly, the performance gap between the per-iteration MSE minimization and the fixed power control is observed to increase with $K$ increasing, which validates the effectiveness on the aggregation error minimization.

%\begin{figure}
%\centering
% %\epsfxsize=1\linewidth
% \setlength{\abovecaptionskip}{-1mm}
%\setlength{\belowcaptionskip}{-1mm}
%    \includegraphics[width=8cm]{Convergence.eps}
%\caption{Convergence analysis of the proposed iterating-based optimization algorithm.}
% \label{fig:Conver}
%\vspace{-0.3cm}
%\end{figure}
%
%Furthermore, the convergence performance of proposed alternating-optimization based algorithm in one channel realization is depicted for the setting of $K =10$ and  $T=100$ in Fig.~\ref{fig:Conver}.
%The values of effective optimality gap are monotonically non-increasing over iterations and converges eventually, thus showing the effectiveness of Algorithm 2.

%\subsection{Air-FedAvg for Handwritten Digit Recognition}\label{Sec_CNN}

\begin{figure*}[htbp]  \vspace{-0.05cm}
  \centering
  \subfigure[Loss value versus $T$ under $K=10$ with i.i.d. data distribution.]
  {\label{CNN_T_fig:OG}\includegraphics[width=8.8cm]{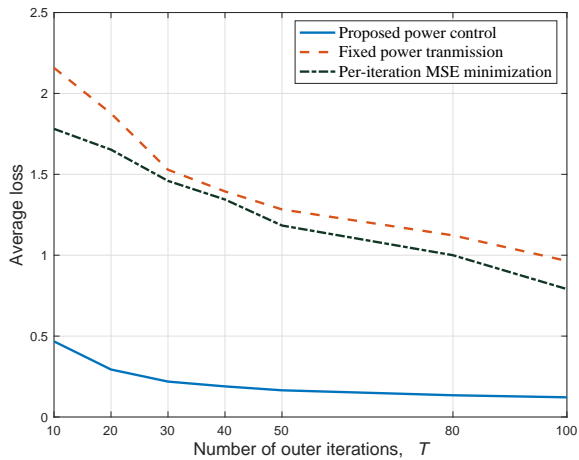}}
%  \vspace{0.08in}
  \subfigure[Test accuracy versus $T$ under $K=10$ with i.i.d. data distribution.]
  {\label{CNN_T_fig:PE}
\includegraphics[width=8.8cm]{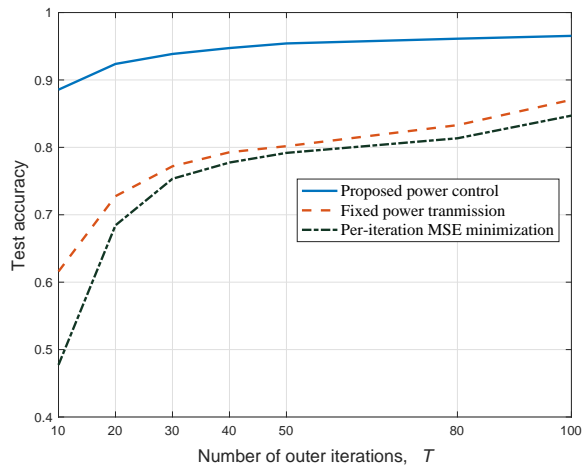}}
  \subfigure[Test accuracy versus $T$ under $K=10$ with non-i.i.d. data distribution.]
  {\label{CNN_T_fig:nonOG}
\includegraphics[width=8.8cm]{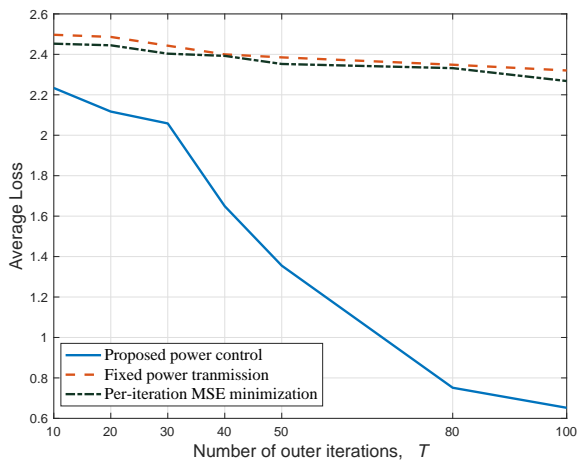}}
  \subfigure[Test accuracy versus $T$ under $K=10$ with non-i.i.d. data distribution.]
  {\label{CNN_T_fig:nonPE}
\includegraphics[width=8.8cm]{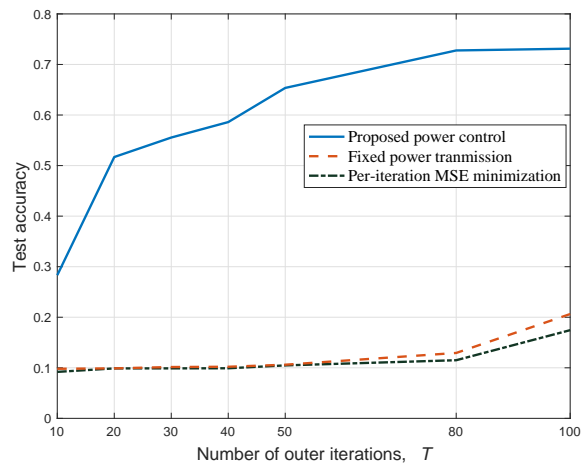}}
  \caption{Learning performance of  Air-FedAvg on MNIST dataset over number of outer iterations.}
  \label{T_fig_CNN}
\vspace{-0.3cm}
\end{figure*}

Next, consider the learning task of handwritten digit recognition using the well-known MNIST datasets, which consists of 10 classes of black-and-white digits ranging from ``0" to ``9". %each image has 784 pixels and a label.
We implement  a 6-layer {\it convolution neural network} (CNN) as the classifier model, which  consists of two $5\times 5$ convolution layers with ReLU activation (the first with 32 channels and the second with 64), each followed by a $2 \times 2$ max pooling; a fully connected layer with 512 units and ReLU activation; and a final softmax output layer ($582,026$ parameter in total). %The local batch size at each edge device is set to be $m_b=512$.
Notice that  Assumptions \ref{Assump_Smooth} and \ref{Assump_PL} may not hold in this case, but our proposed power control policy still work well as will be shown shortly.

Fig.~\ref{T_fig_CNN} shows the learning performance versus the number of outer iterations $T$ with $K=10$ under both i.i.d. and non-i.i.d. data distribution, where the optimized number of local update epochs is $\Omega^{\star}=10$.
%where the learning rate at Figs.~\ref{CNN_fig:OGv_N} and~\ref{CNN_fig:PE_v_N} is set to be diminishing and that in Figs.~\ref{CNN_fig:OG_v_N} and~\ref{CNN_fig:PE_v_N} is fixed to be $0.01$.
It is observed that the proposed power control policy achieves lower loss function values and higher test accuracy than both the fixed power transmission and per-iteration MSE minimization schemes for both the  i.i.d. and non-i.i.d. data.
% These observations are generally consistent with those in Fig. \ref{Fig:Learning_v_N} with the ridge regression model.
By comparing Figs. \ref{CNN_T_fig:nonOG} and \ref{CNN_T_fig:nonPE}  for non-i.i.d. data versus Figs.  \ref{CNN_T_fig:OG} and \ref{CNN_T_fig:PE} for i.i.d. ones, it is observed that the learning performance (in terms of average loss and test accuracy) degrades for non-i.i.d. data, as compared to that for i.i.d. ones. It is also observed that in the case with non-i.i.d. data, the performance gap achieved by the proposed transmission power control designs over benchmarks becomes more significant than that in the case with i.i.d. data.
%Besides, due to the data heterogeneity, the performance would be degraded under the non- i.i.d. setting in Figs. \ref{CNN_T_fig:nonOG} and \ref{CNN_T_fig:nonPE} compared with the i.i.d. one in in Figs. \ref{CNN_T_fig:OG} and \ref{CNN_T_fig:PE}.

%{\color{red}
\subsection{Training Latency Performance Comparison Between Air-FedAvg versus OMA-FedAvg}
\begin{figure*}[htbp] %\vspace{-0.05cm}
  \centering
  \subfigure[Average per-iteration latency (including the communication and computation time) versus $K$.]
  {\label{Latency_fig:PerRound}\includegraphics[width=8.8cm]{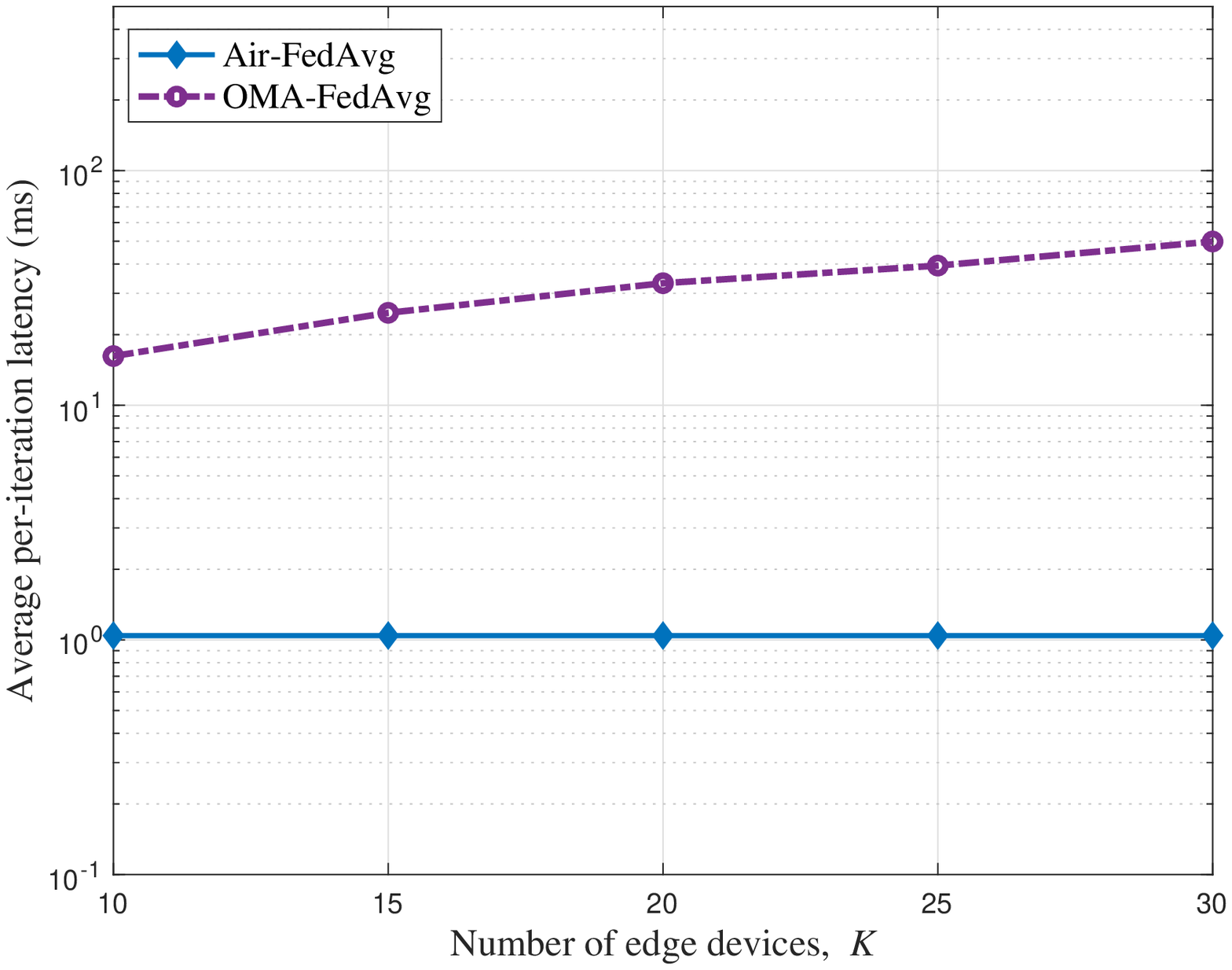}}
%   \vspace{-0.05in}
  \subfigure[Minimum number of required outer iterations versus $K$.]
  {\label{Latency_fig:Round}
\includegraphics[width=8.8cm]{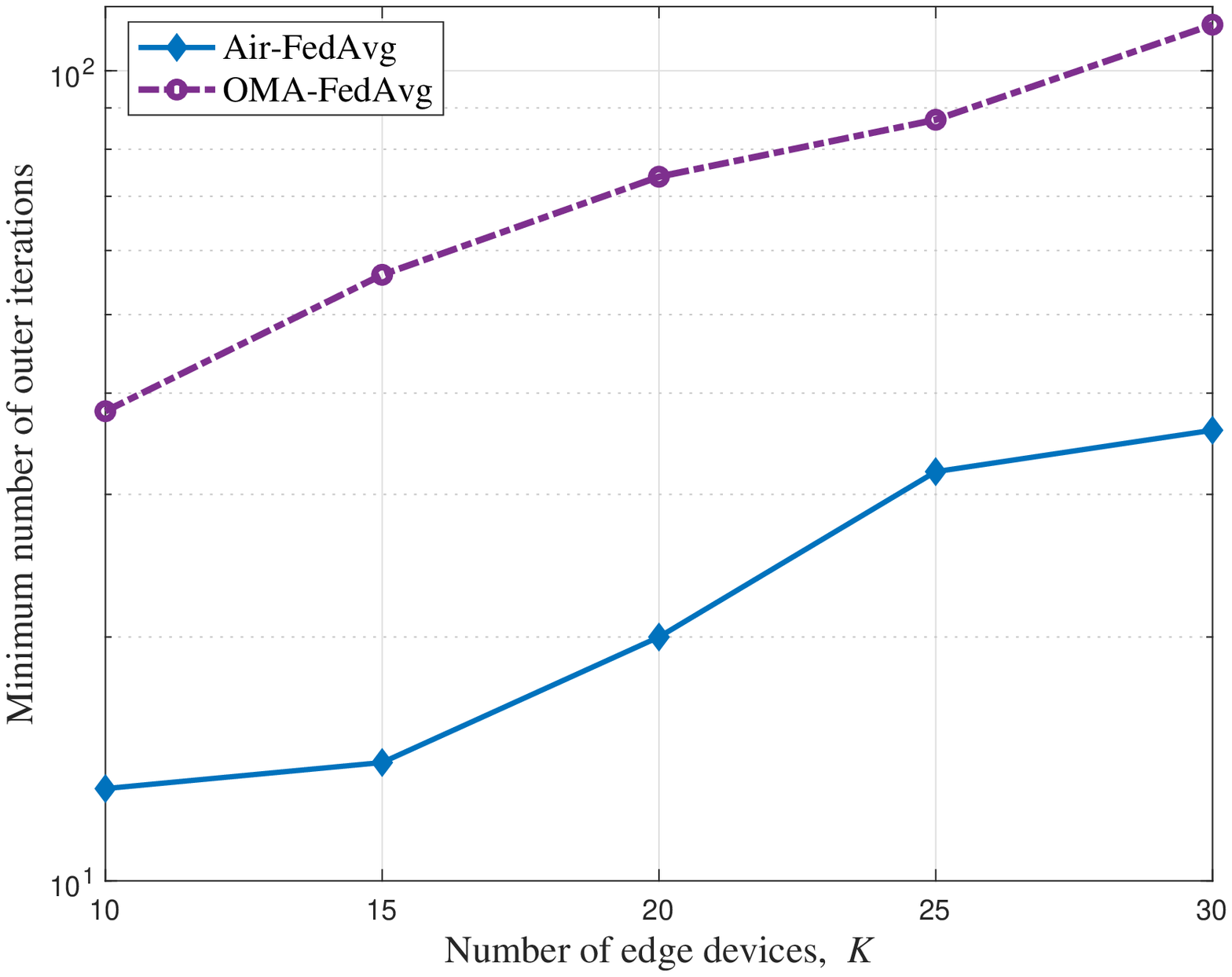}}
%    \vspace{-0.05in}
  \subfigure[Total training time versus $K$ under different required $T$.]
  {\label{Latency_fig:Time}\includegraphics[width=8.8cm]{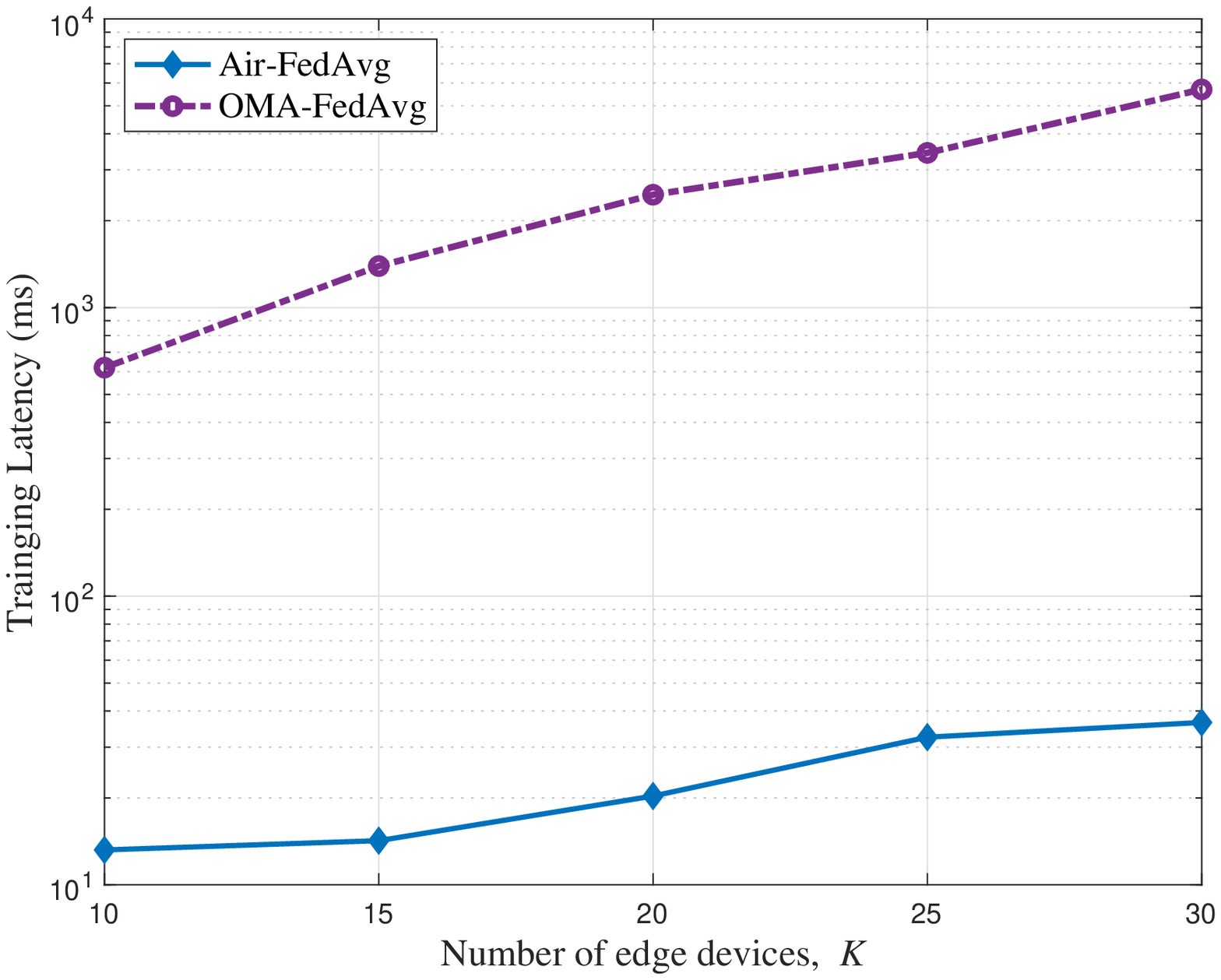}}
%   \vspace{-0.05in}
  \subfigure[Optimality gap versus $K$ under different required $T$.]
  {\label{Latency_fig:OG}
\includegraphics[width=8.8cm]{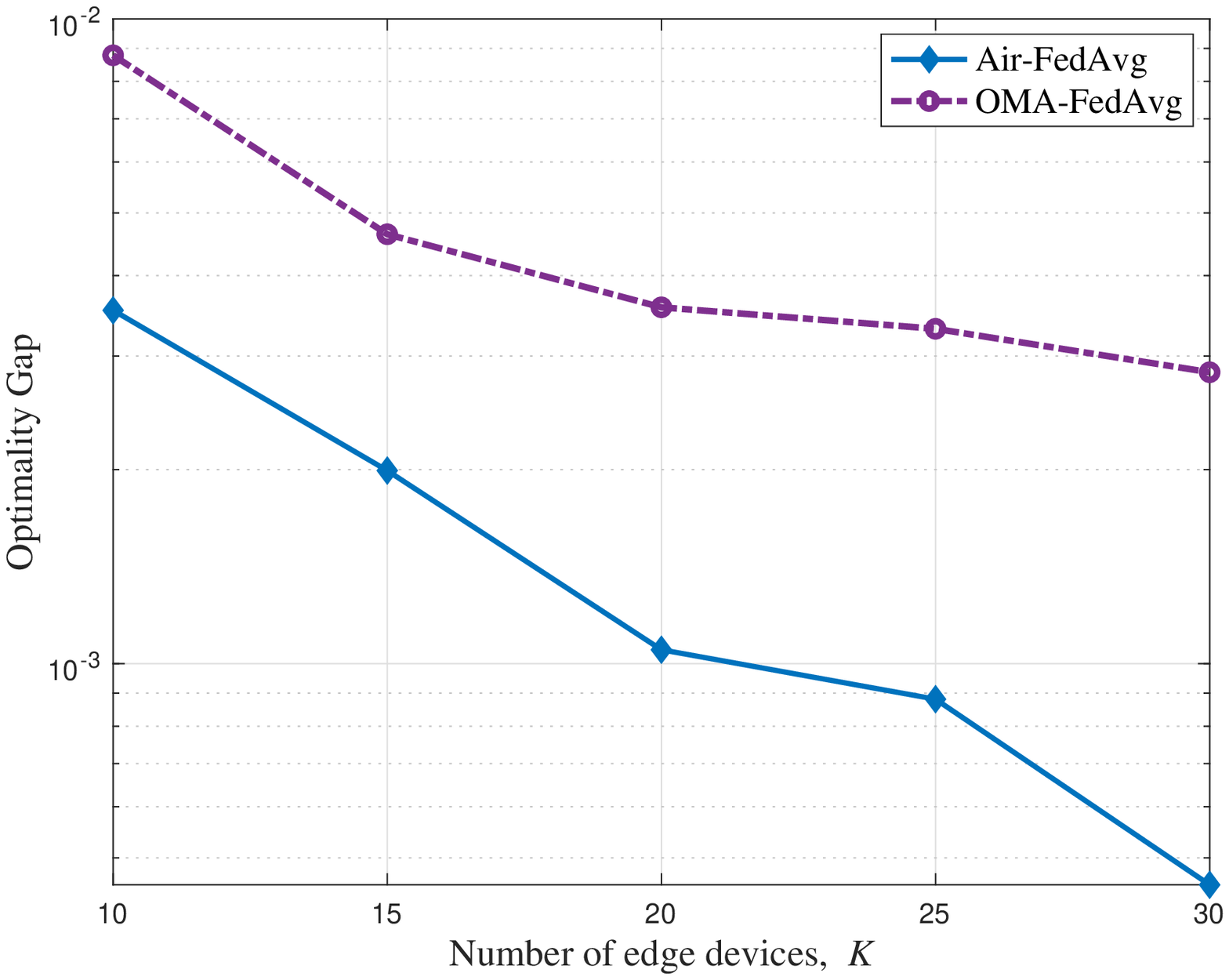}}
  \caption{Comparison of the achieved training latency for Air-FedAvg versus OMA-FedAvg.}
  \label{Latency_fig}
\vspace{-0.4cm}
\end{figure*}
This subsection compares the training latency performance between Air-FedAvg and OMA-FedAvg, by considering the ridge regression. We set the quantization level  as $s=10$ and $S_0=64$ bits for OMA-FedAvg, as well as the bandwidth $B=1$ MHz. The number of CPU cycles required for computing each sample is $c=3000$ cycles/s and the CPU frequency is $f=5$ GHz.  Hence, the local computation latency for each local training update is $\tau_{\rm comp}=c n_b/{f} =0.03$ ms.

%{\color{blue}
Fig.~\ref{Latency_fig} depicts the training latency achieved by Air-FedAvg and OMA-FedAvg versus the number of edge devices $K$, where the optimized numbers of local training epochs for Air-FedAvg and OMA-FedAvg are $5$ and $4$, respectively.
From Fig.~\ref{Latency_fig:PerRound}, it is observed that the average per-iteration latency for OMA-FedAvg increases as $K$ increases, while that for Air-FedAvg keeps unchanged. This is consistent with the discussion in Section \ref{AirVsOMA}, and shows that in this case, the Air-FedAvg reduces the training latency more than 10 times.
Due to the proper power control to combat the model aggregation error, the required number of outer iterations for Air-FedAvg is lower that for OMA-FedAvg as shown in Fig.~\ref{Latency_fig:Round}.
Hence, with the superiority of AirComp in per-iteration latency and potentially also in the required number of outer iterations, the total training latency of Air-FedAvg is significantly lower than that of OMA-FedAvg as observed in Fig.~\ref{Latency_fig:Time}. Besides, Fig.~\ref{Latency_fig:OG} shows that the Air-FedAvg also achieves lower optimality gap than the OMA-FedAvg, thus validating the benefit of AirComp for FEEL like FedAvg.
%}

\section{Conclusion}

This paper studied a new analog NOMA approach named AirComp to facilitate the model-update aggregation for FedAvg at the network edge. For the so-called Air-FedAvg, we first derived a theoretic upper bound for the optimality gap in the presence of over-the-air aggregation errors. Then, based on the derived results, we optimized the transmission power control over outer iterations to minimize the optimality gap, subject to a series of individual power constraints. After that, we investigated the minimum training latency of Air-FedAvg under a targeted optimality gap requirement, and compared the training latency performance versus that of OMA-FedAvg.
Numerical results validated the learning performance gain in Air-FedAvg, achieved by power control optimization, and also showed the benefit of AirComp over OMA in training latency reduction for FedAvg.
This paper revealed that for emerging FEEL applications with coupled communication and computation, new AirComp-enabled multiple access is a promising technique, by exploiting the integrated communication and communication feature over wireless channels.
%techniques with integrated communication and communication such as such as AirComp are emerging to enhance the system performance.
How to extend Air-FEEL and Air-FedAvg designs into other scenarios (e.g., with multi-antenna edge devices/server, assisted by {\it intelligent reflecting surfaces} (IRSs), or powered by renewable energy) are interesting topics worth pursuing in the future.
%{\color{blue}Besides, it is an interesting problem to fairly and quantitatively compare the performance of Air-FEEL versus the  traditional NOMA based FEEL   in terms of the learning performance.
% Generally speaking, the proposed AirComp based solution can achieve significant transmission latency reduction, especially in the large-scale network setting with many devices.
%Although the NOMA can achieve a higher rate with reduced transmission delay,  an additional decoding delay due to the SIC process would be induced unavoidably, thus marginalizing the gain of NOMA in FEEL.
%In this case, how to determine proper approximate optimality gaps for both Air-FEEL and traditional NOMA based FEEL to enable their fair training performance comparison is challenging in practice.
%}

\appendix

\subsection{Proof of Theorem~\ref{Theo_OG_NOMA}}\label{Proof_Theo_OG_NOMA}
To proceed with, we first introduce an auxiliary variable ${\bf g}_t$  as the {\it accumulated gradient} on the server according to SGD, given by
\begin{align}\label{Definition_g}
	{\bf g}_t&=\frac{1}{K}\sum\limits_{k\in\mathcal{K}}  \left( {\bf v}_{t}- {\bf w}_{k,t+1}\right)\notag\\
	&=\frac{1}{K}\sum\limits_{k\in\mathcal{K}}  \left(\sum_{\ell=0 }^{\Omega-1}\gamma_t\nabla \tilde{F}_k({\bf w}_{k,t,\ell},\tilde{\mathcal D}_k)  \right),
\end{align}
based on which we further reformulate the exact aggregated model parameter as
\begin{align}\label{unbiased_v}
	\tilde{\bf v}_{t+1}&={\bf v}_{t}-{\bf g}_{t}={\bf v}_{t}-\frac{1}{K}\sum\limits_{k\in\mathcal{K}}  \left({\bf v}_{t}- {\bf w}_{k,t+1}\right)\notag\\
	&={\bf v}_{t}-\frac{1}{K}\sum\limits_{k\in\mathcal{K}}  \left(\sum_{\ell=0 }^{\Omega-1}\gamma_t\nabla \tilde{F}_k({\bf w}_{k,t,\ell},\tilde{\mathcal D}_k)  \right).
\end{align}
Recall that $\gamma_t $ is the learning rate.
Thus, with the erroneous communication in \eqref{sys_Err}, the update of learning model at outer iteration $t+1$ is represented as
\begin{align}\label{Error_ModelUpdate}
{\bf v}_{t+1}&=\tilde{\bf v}_{t+1}+ {\bm \varepsilon}_{t+1}\notag\\
\!&\!=\!{\bf v}_{t}\!-\!\frac{1}{K}\sum\limits_{k\in\mathcal{K}}\!  \left(\sum_{\ell=0 }^{\Omega-1}\gamma_t\nabla \tilde{F}_k({\bf w}_{k,t,\ell},\tilde{\mathcal D}_k)  \right)+ {\bm \varepsilon}_{t+1}.\!
\end{align}
Next, the proof follows by relating the norm of the gradient to the expected improvement made at each outer iteration.
We start with considering \eqref{Error_ModelUpdate}, based on which it follows that
\begin{align}
\!\!\!\!&F\left({\bf v}_{t+1}\right)- F\left({\bf v}_{t}\right)\notag\\
\!\!\!\! 	&\leq   \left(\nabla F\left({\bf v}_{t}\right)\right)^{\dagger}({\bf v}_{t+1}- {\bf v}_{t}) + \frac{L}{2}\|{\bf v}_{t+1}-{\bf v}_{t}\|^2\notag\\
\!\!\! 	&= \left(\nabla F\left({\bf v}_{t}\right)\right)^{\dagger}({\bf v}_{t}-{\bf g}_{t}+ {\bm \varepsilon}_{t+1}- {\bf v}_{t}) \notag\\
&~~+ \frac{L}{2}\left\|{\bf v}_{t}-{\bf g}_{t}+ {\bm \varepsilon}_{t+1}- {\bf v}_{t}\right\|_2^2\notag\\
%	\!\!\! &= \left(\nabla F\left({\bf v}_{t}\right)\right)^{\dagger}\left({\bm \varepsilon}_{t+1}-{\bf g}_{t}\right)+ \frac{L}{2}\left\|{\bm \varepsilon}_{t+1}-{\bf g}_{t}\right\|_2^2\notag\\
\!\!\!\!\! &\!\!= \!\left(\!\nabla \!F\!\left({\bf v}_{t}\right)\!\right)^{\dagger}\!\!\left(\!\!{\bm \varepsilon}_{t+1}\!\!-\!\frac{1}{K}\!\sum\limits_{k\in\mathcal{K}}\!\!  \left(\sum_{\ell=0 }^{\Omega-1}\!\gamma_t\!\nabla \tilde{F}_k({\bf w}_{k,t,\ell},\!\tilde{\mathcal D}_k)\!\! \right)\!\right)\notag\\
&~+ \frac{L}{2}\!\left\|{\bm \varepsilon}_{t+1}\!-\!\!\frac{1}{K}\!\sum\limits_{k\in\mathcal{K}}\!  \!\left(\sum_{\ell=0 }^{\Omega-1}\!\gamma_t\!\nabla \tilde{F}_k({\bf w}_{k,t,\ell},\!\tilde{\mathcal D}_k) \! \right)\! \right\|^2\!\!\!.	\!\! \!\label{Proof_Smoth1}
\end{align}
%where the above inequality follows Assumption~\ref{Assump_Smooth}  and \label{Definition_g1}.
By taking the statistical expectation at both sides of \eqref{Proof_Smoth1}, we have
\begin{align}
&\mathbb{E}\left(F\left({\bf v}_{t+1}\right)-F\left({\bf v}_{t}\right) \right)\notag\\
&\!\leq\!\left(\nabla F\left({\bf v}_{t}\right)\!\right)^{\dagger}\!\mathbb{E}\!\!\left[\!{\bm \varepsilon}_{t+1}\!\!-\!\!\frac{1}{K}\sum\limits_{k\in\mathcal{K}} \!\! \left(\sum_{\ell=0 }^{\Omega-1}\!\gamma_t\nabla \tilde{F}_k({\bf w}_{k,t,\ell},\!\tilde{\mathcal D}_k)\!\!  \right)\!\right]\notag\\
&~~+\! \frac{L}{2}\!\mathbb{E}\!\left\|{\bm \varepsilon}_{t+1}\!-\!\frac{1}{K}\!\sum\limits_{k\in\mathcal{K}}\!\!  \left(\sum_{\ell=0 }^{\Omega-1}\gamma_t\nabla \tilde{F}_k({\bf w}_{k,t,\ell},\!\tilde{\mathcal D}_k) \!\! \right) \!\right\|^2\notag\\
%	&=\left(\nabla F\left({\bf v}_{t}\right)\right)^{\dagger}\mathbb{E}\left({\bm \varepsilon}_{t+1}\right)-\left(\nabla F\left({\bf v}_{t}\right)\right)^{\dagger}\mathbb{E}\left[\frac{1}{K}\sum\limits_{k\in\mathcal{K}}  \left(\sum_{\ell=0 }^{\Omega-1}\gamma_t\nabla \tilde{F}_k({\bf w}_{k,t,\ell},\tilde{\mathcal D}_k)  \right)\right]\notag\\
%	&~~~+ \frac{L}{2}\mathbb{E}\left\|{\bm \varepsilon}_{t+1}-\frac{1}{K}\sum\limits_{k\in\mathcal{K}}  \left(\sum_{\ell=0 }^{\Omega-1}\gamma_t\nabla \tilde{F}_k({\bf w}_{k,t,\ell},\tilde{\mathcal D}_k)  \right) \right\|_2^2\notag\\
%	&=\left(\nabla F\left({\bf v}_{t}\right)\right)^{\dagger}\mathbb{E}\left[{\bm \varepsilon}_{t+1}\right)-\gamma_t\frac{1}{K}\sum\limits_{k\in\mathcal{K}} \sum_{\ell=0 }^{\Omega-1}  \left(\mathbb{E}\left[\left(\nabla F\left({\bf v}_{t}\right)\right)^{\dagger}\nabla \tilde{F}_k({\bf w}_{k,t,\ell},\tilde{\mathcal D}_k)  \right)\right)+ \frac{L}{2}\mathbb{E}\left\|{\bm \varepsilon}_{t+1}\right\|^2\notag\\
%	&~~~+\frac{L\gamma_t^2}{2}\mathbb{E}\left\|\frac{1}{K}\sum\limits_{k\in\mathcal{K}}  \left(\sum_{\ell=0 }^{\Omega-1}\nabla \tilde{F}_k({\bf w}_{k,t,\ell},\tilde{\mathcal D}_k)  \right) \right\|_2^2- L\gamma_t\mathbb{E}\left[ \left({\bm \varepsilon}_{t+1}\right)^T \left(\frac{1}{K}\sum\limits_{k\in\mathcal{K}}  \left(\sum_{\ell=0 }^{\Omega-1}\nabla \tilde{F}_k({\bf w}_{k,t,\ell},\tilde{\mathcal D}_k)  \right) \right)\right]\notag\\
&=\left(\nabla F\left({\bf v}_{t}\right)\right)^{\dagger}\mathbb{E}\left({\bm \varepsilon}_{t+1}\right)+ \frac{L}{2}\mathbb{E}\left\|{\bm \varepsilon}_{t+1}\right\|^2\notag\\
&~~-\gamma_t\frac{1}{K}\sum\limits_{k\in\mathcal{K}} \sum_{\ell=0 }^{\Omega-1}  \mathbb{E}\left(\left(\nabla F\left({\bf v}_{t}\right)\right)^{\dagger}\nabla \tilde{F}_k({\bf w}_{k,t,\ell},\tilde{\mathcal D}_k) \right)\notag\\
	&~~+\frac{L\gamma_t^2}{2}\mathbb{E}\left\|\frac{1}{K}\sum\limits_{k\in\mathcal{K}}  \left(\sum_{\ell=0 }^{\Omega-1}\nabla \tilde{F}_k({\bf w}_{k,t,\ell},\tilde{\mathcal D}_k)  \right) \right\|^2\notag\\
	&~~- \!L\gamma_t\frac{1}{K}\sum\limits_{k\in\mathcal{K}} \sum_{\ell=0 }^{\Omega-1}\mathbb{E}\left( \left({\bm \varepsilon}_{t+1}\right)^T \nabla \tilde{F}_k({\bf w}_{k,t,\ell},\!\tilde{\mathcal D}_k)\!  \right).\label{Proof1_GraBia}
	\end{align}
%where the denominator $m_b$ in \eqref{Proof1_eq1} is induced from Assumption \ref{Assum_VarianceBound} and Equation \eqref{sys_LocalGradient}.
By applying the inequality of arithmetic and geometric means, i.e., ${\bm a_1}^T{\bm a_2}\leq \frac{x\|{\bm a_1}\|^2 }{2}+\frac{\|{\bm a_2}\|^2 }{2x}$,  it then follows  from \eqref{Proof1_GraBia} that
\begin{align}
&\mathbb{E}\left(F\left({\bf v}_{t+1}\right)-F\left({\bf v}_{t}\right) \right)\notag\\
&\leq\frac{\gamma_t\left\|\nabla F\left({\bf v}_{t}\right)\right\|^2}{2} \!+\!\frac{\left\|\mathbb{E}\left({\bm \varepsilon}_{t+1}\right)\right\|^2}{2\gamma_t}
\!+ \!\!\frac{L}{2}\mathbb{E}\left\|{\bm \varepsilon}_{t+1}\right\|^2\notag\\
&~-\gamma_t\frac{1}{K}\sum\limits_{k\in\mathcal{K}} \sum_{\ell=0 }^{\Omega-1}  \mathbb{E}\left[\left(\nabla F\left({\bf v}_{t}\right)\right)^{\dagger}\nabla \tilde{F}_k({\bf w}_{k,t,\ell},\tilde{\mathcal D}_k) \right]\!\notag\\
	&~+\frac{L\gamma_t^2}{2}\mathbb{E}\left\|\frac{1}{K}\sum\limits_{k\in\mathcal{K}}  \left(\sum_{\ell=0 }^{\Omega-1}\nabla \tilde{F}_k({\bf w}_{k,t,\ell},\tilde{\mathcal D}_k)  \right) \right\|^2\notag\\
	&~- L\gamma_t\frac{1}{K}\sum\limits_{k\in\mathcal{K}} \sum_{\ell=0 }^{\Omega-1}\mathbb{E}\left[ \left({\bm \varepsilon}_{t+1}\right)^T \nabla \tilde{F}_k({\bf w}_{k,t,\ell},\tilde{\mathcal D}_k)   \right]\notag\\
	&\leq\frac{\gamma_t\left\|\nabla F\left({\bf v}_{t}\right)\right\|^2}{2} +\frac{\left\|\mathbb{E}\left({\bm \varepsilon}_{t+1}\right)\right\|^2}{2\gamma_t}
+ \frac{L}{2}\mathbb{E}\left\|{\bm \varepsilon}_{t+1}\right\|^2\notag\\
&~-\gamma_t\frac{1}{K}\sum\limits_{k\in\mathcal{K}} \sum_{\ell=0 }^{\Omega-1}  \mathbb{E}\left[\left(\nabla F\left({\bf v}_{t}\right)\right)^{\dagger}\nabla \tilde{F}_k({\bf w}_{k,t,\ell},\tilde{\mathcal D}_k) \right]\notag\\
	&~+\!\frac{L\gamma_t^2}{2}\mathbb{E}\left\|\frac{1}{K}\sum\limits_{k\in\mathcal{K}} \! \left(\sum_{\ell=0 }^{\Omega-1}\nabla \tilde{F}_k({\bf w}_{k,t,\ell},\tilde{\mathcal D}_k) \! \!\right) \!\right\|^2\notag\\
	&~+ L\gamma_t\frac{1}{K}\sum\limits_{k\in\mathcal{K}}\! \sum_{\ell=0 }^{\Omega-1}\mathbb{E}\!\left(\!\frac{L\left\|{\bm \varepsilon}_{t+1}\right\|^2}{2} +\frac{\left\|\nabla \tilde{F}_k({\bf w}_{k,t,\ell},\!\tilde{\mathcal D}_k)\right\|^2}{2L}\!\right)\!\notag
		\end{align}
		\begin{align}
	&= \frac{\gamma_t\left\|\nabla F\left({\bf v}_{t}\right)\right\|^2}{2} +\frac{\left\|\mathbb{E}\left({\bm \varepsilon}_{t+1}\right)\right\|^2}{2\gamma_t}
 + \!\frac{L^2\gamma_t \Omega}{2}\mathbb{E}\left\|{\bm \varepsilon}_{t+1}\right\|^2\notag\\
&~-\gamma_t\frac{1}{K}\sum\limits_{k\in\mathcal{K}} \sum_{\ell=0 }^{\Omega-1}  \mathbb{E}\left[\left(\nabla F\left({\bf v}_{t}\right)\right)^{\dagger}\nabla \tilde{F}_k({\bf w}_{k,t,\ell},\!\tilde{\mathcal D}_k) \right]\notag\\
	&~+\!\frac{L\gamma_t^2}{2}\mathbb{E}\!\left\|\frac{1}{K}\!\sum\limits_{k\in\mathcal{K}} \!\! \left(\sum_{\ell=0 }^{\Omega-1}\nabla \tilde{F}_k({\bf w}_{k,t,\ell},\tilde{\mathcal D}_k) \!\! \right) \!\right\|^2+ \frac{L}{2}\mathbb{E}\left\|{\bm \varepsilon}_{t+1}\right\|^2\!\!\!\notag\\
&	~+\frac{\gamma_t}{2}\frac{1}{K}\sum\limits_{k\in\mathcal{K}} \sum_{\ell=0 }^{\Omega-1}\mathbb{E}\left\|\nabla \tilde{F}_k({\bf w}_{k,t,\ell},\!\tilde{\mathcal D}_k)\right\|^2\notag\!\!.
	\end{align}
%	where \eqref{Proof1_equ_Dk} follows that $\frac{1}{K}\sum\limits_{k\in\mathcal{K}} =1$.
By further applying the equality ${\bm a_1}^T{\bm a_2}=\frac{1}{2}\left(\|{\bm a_1}\|^2+\|{\bm a_2}\|^2-\|{\bm a_1}-{\bm a_2}\|^2 \right)$,  it follows from the above inequality that
\begin{align}
\!	\!&\mathbb{E}\!\left(F\left({\bf v}_{t+1}\right)\!-\!F\left({\bf v}_{t}\right) \right)\leq\!\frac{\gamma_t\!\left\|\nabla F\left({\bf v}_{t}\right)\right\|^2}{2} \!+\!\frac{\left\|\mathbb{E}\!\left({\bm \varepsilon}_{t+1}\right)\right\|^2}{2\gamma_t}\notag\\
&~~+ \!\frac{L\!+\!L^2\gamma_t \Omega}{2}\mathbb{E}\!\left(\left\|{\bm \varepsilon}_{t+1}\right\|^2\right)\!
	-\!\frac{\gamma_t \Omega\!\left\|\nabla F\left({\bf v}_{t}\right)\right\|^2}{2}\notag\\
	\!\!&~~+\!\frac{L\gamma_t^2}{2}\mathbb{E}\left\|\frac{1}{K}\!\sum\limits_{k\in\mathcal{K}}\!\!  \left(\sum_{\ell=0 }^{\Omega-1}\nabla \tilde{F}_k({\bf w}_{k,t,\ell},\!\tilde{\mathcal D}_k) \! \right) \!\right\|^2\!\!\notag\\
	&~~+\frac{\gamma_t}{2}\frac{1}{K}\!\sum\limits_{k\in\mathcal{K}}\! \sum_{\ell=0 }^{\Omega-1}  \mathbb{E}\left\|\nabla F\left({\bf v}_{t}\right)- \!\nabla \tilde{F}_k({\bf w}_{k,t,\ell},\!\tilde{\mathcal D}_k) \right\|^2.\label{Proof1_Func1}\!
	\end{align}
To proceed with, it would be necessary to prove the following lemmas.
\begin{lemma}[Bounding the gradients of ${\bf v}_{t}$ and local gradient from randomly local update]\label{Lemma_vt} \emph{With Assumptions \ref{Assump_Smooth}, \ref{Assum_VarianceBound} and \ref{Assum_GraDiv}, and we have
\begin{align}
\!\!\!	&\!\!\mathbb{E}\!\left\|\nabla\! F\left({\bf v}_{t}\right)\!-\! \nabla \!\tilde{F}_k({\bf w}_{k,t,\ell},\!\tilde{\mathcal D}_k) \right\|^2\!\!\leq\! \delta_k^2\! +\!\phi_k^2 +\!L^2\gamma_t^2 \Omega^2 G_k^2.
\end{align}
}	
\end{lemma}
\begin{IEEEproof}
Please refer to Appendix~\ref{Proof_Lemma_vt}.
\end{IEEEproof}
	
%\begin{lemma}[Bounding the accumulated local gradients]\label{Lemma_gt} \emph{

From Assumption \ref{Assum_GraBound}, the upper bound of the accumulated local gradients is given by
\begin{align}
	&\mathbb{E}\left\|\frac{1}{K}\sum\limits_{k\in\mathcal{K}}  \left(\sum_{\ell=0 }^{\Omega-1}\nabla \tilde{F}_k({\bf w}_{k,t,\ell},\tilde{\mathcal D}_k)  \right) \right\|^2\notag\\
	&\leq \frac{1}{K}\sum\limits_{k\in\mathcal{K}}  \mathbb{E}\left\|\sum_{\ell=0 }^{\Omega-1}\nabla \tilde{F}_k({\bf w}_{k,t,\ell},\tilde{\mathcal D}_k)   \right\|^2\notag\\
	&\leq  \frac{\Omega}{K}\sum\limits_{k\in\mathcal{K}}  \sum_{\ell=0 }^{\Omega-1}\mathbb{E}\left\|\nabla \tilde{F}_k({\bf w}_{k,t,\ell},\tilde{\mathcal D}_k)   \right\|^2\leq \frac{\Omega^2}{K}\sum\limits_{k\in\mathcal{K}}  G_k^2,\label{Accumulated_local_gradients}
\end{align}
where the first inequality is derived from the Jensen's inequality.
Then, combining the Lemma \ref{Lemma_vt} and inequality \eqref{Accumulated_local_gradients}, the inequality \eqref{Proof1_Func1} can be further relaxed into
\begin{align}
&\mathbb{E}\left(F\left({\bf v}_{t+1}\right)-F\left({\bf v}_{t}\right) \right)\notag\\
&\leq-\frac{(\Omega-1)\gamma_t\left\|\nabla F\left({\bf v}_{t}\right)\right\|^2}{2} +\frac{\left\|\mathbb{E}\left({\bm \varepsilon}_{t+1}\right)\right\|^2}{2\gamma_t}\notag\\
&~~~+\frac{\gamma_t}{2}\frac{1}{K}\sum\limits_{k\in\mathcal{K}} \sum_{\ell=0 }^{\Omega-1}  \left(\delta_k^2 +\phi_k^2 +L^2\gamma_t^2 \Omega^2 G_k^2 \right)\notag\\
\!	&~~~+ \frac{L+L^2\gamma_t \Omega}{2}\mathbb{E}\left\|{\bm \varepsilon}_{t+1}\right\|^2+\frac{L\gamma_t^2\Omega^2}{2K}\sum\limits_{k\in\mathcal{K}}  G_k^2\notag\\
	 &=-\frac{(\Omega-1)\gamma_t\left\|\nabla F\left({\bf v}_{t}\right)\right\|^2}{2} +\frac{\left\|\mathbb{E}\left({\bm \varepsilon}_{t+1}\right)\right\|^2}{2\gamma_t}\notag\\
&~~~+ \gamma_t \Omega\frac{1}{K}\sum\limits_{k\in\mathcal{K}} \frac{\delta_k^2 +\phi_k^2}{2}+\frac{L\!+\!L^2\gamma_t \Omega}{2}\mathbb{E}\left\|{\bm \varepsilon}_{t+1}\right\|^2\notag\\
&~~~+\frac{1}{K}\left(\frac{L^2\gamma_t^3E^3 }{2}  +\frac{L\gamma_t^2\Omega^2}{2}\right)   \sum\limits_{k\in\mathcal{K}}G_k^2\notag\\
\!\!&\!\leq\!-\frac{(\Omega\!-\!1)\gamma_t\!\left\|\nabla \!F\!\left({\bf v}_{t}\right)\right\|^2}{2} \!+\!\frac{\left\|\mathbb{E}\left({\bm \varepsilon}_{t+1}\!\right)\right\|^2}{2\gamma_t}\!\notag\\
&~~~+\frac{\gamma_t \Omega}{K}\!\sum\limits_{k\in\mathcal{K}}\! \frac{\delta_k^2 \!+\!\phi_k^2}{2}\!+\!\frac{L\!\!+\!L^2\gamma_t \Omega}{2}\mathbb{E}\left\|{\bm \varepsilon}_{t+1}\right\|^2\!+\!\frac{L\gamma_t^2\Omega^2  }{K}\!\!\sum\limits_{k\in\mathcal{K}}\!G_k^2,\! \notag%\label{Proof1_Func1},
\end{align}
where the last inquality follows from $\gamma_t \Omega\leq 1/L$.
Next, by applying Assumption \ref{Assump_PL},  we have
\begin{align*}
&\mathbb{E}\left[F\left({\bf v}_{t+1}\right)\right]-F^{\star}\leq\left(1-(\Omega-1)\mu\gamma_t\right)\left(\mathbb{E}\left(F\left({\bf v}_{t}\right)\right)-F^{\star}\right)  \notag\\
&~~~+\frac{\left\|\mathbb{E}\left({\bm \varepsilon}_{t+1}\right)\right\|^2}{2\gamma_t}+\frac{L\gamma_t^2\Omega^2  }{K}\sum\limits_{k\in\mathcal{K}}G_k^2\notag\\
&~~~+ \gamma_t \Omega\frac{1}{K}\sum\limits_{k\in\mathcal{K}} \frac{\delta_k^2 +\phi_k^2}{2}+\frac{L+L^2\gamma_t \Omega}{2}\mathbb{E}\left(\left\|{\bm \varepsilon}_{t+1}\right\|^2\right).
\end{align*}

Through some further algebraic manipulation over the above inequality, we have \eqref{OG_NOMA}.
This thus completes the proof.

%{\color{blue}

\subsection{Proof of Lemma~\ref{Lemma_vt}}\label{Proof_Lemma_vt}
%To proceed with, we first prove the upper bound of $\mathbb{E}\left\|\nabla F\left({\bf v}_{t}\right)- \nabla \tilde{F}_k({\bf w}_{k,t,\ell},\tilde{\mathcal D}_k) \right\|^2$.
Recall that ${\bf v}_{t}={\bf w}_{k,t,0}, \forall k\in\mathcal{K}$.
By applying $\|{\bm a_1}-{\bm a_2} \|\leq \|{\bm a_1}\|^2+\|{\bm a_2}\|^2$ and Assumption \ref{Assum_GraDiv}, it follows
\begin{align}\!
	&\mathbb{E}\left\|\nabla F\left({\bf v}_{t}\right)- \nabla \tilde{F}_k({\bf w}_{k,t,\ell},\tilde{\mathcal D}_k) \right\|^2\notag\\
	&=\mathbb{E}\left\|\left(\nabla F\left({\bf v}_{t}\right)-\!\nabla F_k({\bf v}_{t} )\right) \!+\!\left(\!\nabla \!F_k({\bf v}_{t} ) \!-\! \nabla \tilde{F}_k({\bf w}_{k,t,\ell},\tilde{\mathcal D}_k)\!\right) \!\right\|^2\notag\\
	&\leq \mathbb{E}\left\|\nabla F\left({\bf v}_{t}\right)-\nabla F_k({\bf v}_{t})\right\|^2 \notag\\
	&~~~+\mathbb{E}\left\|\nabla F_k({\bf v}_{t}) - \nabla \tilde{F}_k({\bf w}_{k,t,\ell},\tilde{\mathcal D}_k) \right\|^2\notag\\
	&\leq \delta_k^2 +\mathbb{E}\left\|\nabla F_k({\bf w}_{k,t,0} ) - \nabla \tilde{F}_k({\bf w}_{k,t,\ell},\tilde{\mathcal D}_k) \right\|^2\notag\\	
%	&~= \delta_k^2 +\mathbb{E}\left\|\nabla F_k({\bf w}_{k,t,0} ) -\nabla F_k({\bf w}_{k,t,\ell} )+\nabla F_k({\bf w}_{k,t,\ell} )- \nabla \tilde{F}_k({\bf w}_{k,t,\ell},\tilde{\mathcal D}_k) \right\|^2\notag\\	
	&\leq \delta_k^2 +\mathbb{E}\left\|\nabla F_k({\bf w}_{k,t,\ell} )- \nabla \tilde{F}_k({\bf w}_{k,t,\ell},\tilde{\mathcal D}_k) \right\|^2\notag\\
	&~~+\mathbb{E}\left\|\nabla F_k({\bf w}_{k,t,0} ) -\nabla F_k({\bf w}_{k,t,\ell} )\right\|^2.\notag
	\end{align}
	Based on  Assumption \ref{Assum_VarianceBound}, it follows
	\begin{align}
	&\mathbb{E}\left\|\nabla F\left({\bf v}_{t}\right)- \nabla \tilde{F}_k({\bf w}_{k,t,\ell},\tilde{\mathcal D}_k) \right\|^2\notag\\
	&\leq \delta_k^2 +\phi_k^2 +\mathbb{E}\left\|\nabla F_k({\bf w}_{k,t,0} ) -\nabla F_k({\bf w}_{k,t,\ell} )\right\|^2\notag\\	
	&\leq \delta_k^2 +\phi_k^2 +L^2\mathbb{E}\left\|{\bf w}_{k,t,0}  -{\bf w}_{k,t,\ell} \right\|^2\label{Proof1_inequ_model}\\	
%	&~= \delta_k^2 +\phi_k^2 +L^2\mathbb{E}\left\|\sum_{j=0}^{\ell-1}\left({\bf w}^k_{t,j+1}  -{\bf w}^k_{t,j} \right)\right\|^2\notag\\		
	&= \delta_k^2 +\phi_k^2 +L^2\gamma_t^2\mathbb{E}\left\|\sum_{j=0}^{\ell-1}\nabla \tilde{F}_k({\bf w}_{t,j},\tilde{\mathcal D}_k)\right\|^2\notag\\
	&\leq \delta_k^2 +\phi_k^2 +L^2\gamma_t^2\ell \sum_{j=0}^{\ell-1}\mathbb{E}\left\|\nabla \tilde{F}_k({\bf w}_{t,j},\tilde{\mathcal D}_k)\right\|^2\label{Proof1_inequ_gra2} \\	
%	&~\leq \delta_k^2 +\phi_k^2 +L^2\gamma_t^2\ell^2 G_k^2,\label{Proof1_equ_gra2}\\
	&\leq \delta_k^2 +\phi_k^2 +L^2\gamma_t^2 \Omega^2 G_k^2,\label{Proof1_equ_gra2}
	\end{align}
where $\phi_k^2=\frac{\hat{\phi}_k^2}{ n_b}, \forall k\in\mathcal{K}$, %the inequality \eqref{Proof1_equ_gra} follows the Assumption \ref{Assum_VarianceBound},
 the inequality \eqref{Proof1_inequ_model} follows from \eqref{Smooth_gra_k} in Assumption \ref{Assump_Smooth}, the inequality \eqref{Proof1_inequ_gra2} follows from the inequality of arithmetic and geometric means, i.e., $(\sum_{i=1}^I a_i)^2\leq I(\sum_{i=1}^I a_i^2)$, and the inequality \eqref{Proof1_equ_gra2} holds due to Assumption \ref{Assum_VarianceBound}. This thus completes the proof.

\subsection{Proof of Lemma~\ref{lemma_NOMA_power_q}}\label{Proof_lemma_NOMA_power_q}

Notice that problem \eqref{NOMA_power_q} is convex and satisfies the Slater's condition, and thus strong duality holds between problem \eqref{NOMA_power_q} and its Lagrange dual problem \cite{cvx}. Therefore, we apply the Lagrange duality method to optimally solve problem \eqref{NOMA_power_q}.
Let $\lambda_k\ge 0$ denote the dual variable associated with  the $k$-th constraints in \eqref{P1_q_ave1}. Then the partial Lagrangian of problem \eqref{NOMA_power_q} is
\begin{align*}
\mathcal{L}\left(\{\hat{p}_{k,t}\}\right)=&\sum_{i=1}^{T} a_t  \sum\limits_{k\in\mathcal K} c_k \left(\frac{h_{k,t}\hat{p}_{k,t}}{\sqrt{\eta_t^*}}-1\right)^2\notag\\
&+\sum\limits_{k\in\mathcal{K}} \frac{\lambda_k}{T}\left(\sum_{t=1}^{T}\hat{p}_{k,t}^2- \tilde{P}^{\rm ave}_k\right)\notag.
\end{align*}
Then the dual function is
\begin{align}\label{Air_q_DualFucntion}
W_1(\{\lambda_k\})=\min_{\{\hat{p}_{k,t}\ge 0\}} ~&\mathcal{L}\left(\{\hat{p}_{k,t}\}\right)\\
{\rm s.t.}~~&\eqref{P1_q_max1}.\notag
\end{align}
Accordingly, the dual problem of problem (P1) is given as
\begin{align}
\mathbf{D1:} \min_{\{\lambda_k\ge 0\}} ~&W_2(\{\lambda_k\}).
\end{align}
Due to the fact that the strong duality holds between problems \eqref{NOMA_power_q} and (D1), we can solve problem \eqref{NOMA_power_q} by equivalently solving its dual problem (D1). For notational convenience, let $\{\hat{p}_{k,t}^{\rm opt}\}$ denote the optimal primal solution to problem \eqref{NOMA_power_q}, and $\{\lambda_k^{\rm opt}\}$  denote the optimal dual solution to problem (D1). Next, we first evaluate the dual function $W_1(\{\lambda_k\})$ under any given feasible $\{\lambda_k\}$, and then obtain the optimal dual variables $\{\lambda_k^{\rm opt}\}$ to maximize $W_1(\{\lambda_k\})$.

First, we obtain $W_1(\{\lambda_k\})$ by solving problem \eqref{Air_q_DualFucntion} under any given feasible $\{\lambda_k\}$. This can be decomposed into a sequence of subproblems each for optimizing the transmission power scaling factor in edge device $k$ at outer iteration $t$, i.e.,
\begin{align}\label{Air_q_t}
\min_{0\leq \hat{p}_{k,t}\leq\sqrt{\tilde{P}^{\rm max}_k}} ~&a_t c_k \left(\frac{h_{k,t}\hat{p}_{k,t}}{\sqrt{\eta_t}}-1\right)^2+\frac{\lambda_k}{T}\hat{p}_{k,t}^2.
\end{align}
Via the first-order derivation of the objective function in \eqref{Air_q_t}, we have the following lemma.
\begin{lemma}\label{lemma_Air_q_t}\emph{The optimal solution to problem \eqref{Air_q_t} denoted by $\hat{p}_{k,t}^*$ is given as
\begin{align}
\hat{p}_{k,t}^*=\min \left[\frac{a_t c_k h_{k,t}T \sqrt{\eta_t}}{ a_t c_k\left( h_{k,t}\right)^2T + \lambda_k \eta_t},\sqrt{ \tilde{P}^{\rm max}_k}~\right].
\end{align}
}
\end{lemma}
Therefore, with Lemma \ref{lemma_Air_q_t},  problem \eqref{Air_q_t} is solved, and the dual function $W_1(\{\lambda_k\})$ is accordingly obtained. It thus remains to find the optimal $\{\lambda_k\}$.
Since the dual function $W_1(\{\lambda_k\})$ is concave but non-differentiable in general, we use the ellipsoid method \cite{Gradient},	 to obtain the optimal dual variables. Note that for the objective function in \eqref{Air_q_DualFucntion}, the subgradient w.r.t. $\lambda_k, \forall k$, is $ \frac{1}{T}\sum_{t=1}^{T}(\hat{p}_{k,t}^{*})^2- \tilde{P}^{\rm ave}_k $.
By replacing $\{\lambda_k\}$ in Lemma~\ref{lemma_Air_q_t} with the obtained optimal dual variables $\{\lambda_k^{\rm opt}\}$, the optimal solution to problem \eqref{NOMA_power_q} is accordingly obtained as shown in \eqref{Air_q_Opt}. This thus completes the proof.

\bibliography{AirCompforFL}
\bibliographystyle{IEEEtran}

\end{document}